\documentclass[11pt,preprint]{aastex}
\usepackage{graphics}
\usepackage{epsfig}
\usepackage{longtable}
\def\gsim{\lower 2pt \hbox{$\, \buildrel {\scriptstyle >}\over
{\scriptstyle \sim}\,$}}
\def\lsim{\lower 2pt \hbox{$\, \buildrel {\scriptstyle <}\over
{\scriptstyle \sim}\,$}}

\def\HI{H{\small I}\ }

\def\cor{\widehat=}

\shortauthors{}
\shorttitle{}

\begin{document}

\title{X-ray emission from the Sombrero galaxy: a galactic-scale outflow}
\author{Zhiyuan Li\altaffilmark{1}, 
  Christine Jones\altaffilmark{1}, William R.~Forman\altaffilmark{1}, 
   Ralph P.~Kraft\altaffilmark{1}, Dharam V.~Lal\altaffilmark{1}, \\ Rosanne Di Stefano\altaffilmark{1},
  Lee R.~Spitler\altaffilmark{2}, Shikui Tang\altaffilmark{3}, Q.~Daniel
  Wang\altaffilmark{3}, \\
  Marat Gilfanov\altaffilmark{4}, Mikhail Revnivtsev\altaffilmark{5,6}}
\altaffiltext{1}{Harvard-Smithsonian Center for Astrophysics, 60
  Garden Street, Cambridge, MA 02138, USA; zyli@cfa.harvard.edu}
\altaffiltext{2}{Centre for Astrophysics and Supercomputing, Swinburne University, Hawthorn, VIC 3122, Australia}
\altaffiltext{3}{Department of Astronomy, University of Massachusetts,
  710 North Pleasant Street, Amherst, MA 01003, USA}
\altaffiltext{4}{Max-Planck-Institut f${\rm \ddot{u}}$r Astrophysik,
  Karl-Schwarzschild-Str 1, 85741 Garching bei M${\rm \ddot{u}}$nchen, Germany}
\altaffiltext{5}{Excellence Cluster Universe, Technische Universit\"at M\"unchen, Boltzmannstr.2, 85748 Garching, Germany}
\altaffiltext{6}{Space Research Institute, Russian Academy of
  Sciences, Profsoyuznaya 84/32, 117997 Moscow, Russia}

\begin{abstract}
Based on new and archival {\sl Chandra} observations of the Sombrero galaxy (M~104=NGC~4594), we study the X-ray emission from its nucleus and the
extended X-ray emission in and around its massive stellar bulge.
We find that the 0.3-8 keV luminosity of the nucleus appears  
constant at $\sim$$2.4\times10^{40}{\rm~ergs~s^{-1}}$, or $\sim$10$^{-7}$ of
its Eddington luminosity, on three epochs between December 1999
and April 2008, but drops by a factor of two in the November 2008 observation.
The 2-6 keV unresolved
emission from the bulge region closely follows the K-band star light and most likely arises from
unresolved stellar sources.
At lower energies, however, the unresolved emission
reaches a galactocentric radius of at least 23 kpc, significantly beyond the
extent of the star light, clearly indicating the presence
of diffuse hot gas. 
We isolate the emission of the gas by properly accounting for
the emission from unresolved stellar sources, predominantly 
cataclysmic variables and coronally active binaries, whose
quasi-universal X-ray emissivity was recently established.
We find a gas
 temperature of $\sim$0.6 keV with little variation across the field
of view, except for a lower temperature of $\sim$0.3 keV along the
stellar disk. 
The metal abundance is not well constrained due to the limited
counting statistics, but is consistent with metal-enrichment by Type Ia supernovae.
We measure a total intrinsic 0.3-2 keV luminosity of
$\sim$$2{\times}10^{39}{\rm~ergs~s^{-1}}$,  which corresponds to only one
percent of the available energy input by Type Ia supernovae in the
bulge, but is comparable to the prediction by 
the latest galaxy formation models for disk galaxies as massive as Sombrero. 
However, such numerical models do not fully account for internal feedback
processes, such as nuclear feedback and stellar feedback,
against accretion from the intergalactic medium. 
On the other hand, we find no evidence for either the nucleus or the very
modest star-forming activities in the disk to be a dominant heating source
for the diffuse gas. We also show that neither the expected energy released by Type Ia supernovae nor
the expected mass returned by evolved stars is recovered by observations.   
We argue that in Sombrero a galactic-scale subsonic outflow of hot gas 
continuously removes much of the ``missing'' energy and
mass input from the bulge region. The observed density and temperature
distributions of such an
outflow, however, continues to pose challenges to theoretical studies.     
\end{abstract}

\section{Introduction} {\label{sec:intro}}
Early X-ray observations of nearby galactic
spheroids (i.e., elliptical galaxies and bulges of disk galaxies) have
established two primary X-ray emitters: diffuse hot gas and discrete
sources (e.g., Forman, Jones \& Tucker 1985; Van Speybroeck et
al.~1979; Trinchieri \& Fabbiano 1985). 
The hot gas is perhaps ubiquitous in early-type galaxies, but is found to  
dominate the overall X-ray emission only in gas-rich, massive
elliptical galaxies.
In low- and intermediate-mass galaxies, 
the discrete sources, primarily stellar binaries
whose collective emission roughly scales with the host
galaxy's stellar mass (Gilfanov 2004; Revnivtsev et al.~2007), generally overwhelm the hot gas.
An X-ray nucleus, with a high incidence rate in nearby early-type galaxies 
(Zhang et al.~2009; Gallo et al.~2010), may further complicate 
the observed X-ray emission.
Studies of the hot gas in the less massive early-type galaxies thus have been warranted only with
the {\sl Chandra X-ray Observatory} (Weisskopf et al.~2002), owing to its superb angular
resolution for isolating discrete sources 
and its excellent sensitivity for tracing faint, extended X-ray emission.

Many recent efforts to study hot gas in
elliptical galaxies
(e.g., Fukazawa et al.~2006; Diehl \& Statler 2007; Memola et
al.~2009), however, have
not led to a consensus on the physical properties of the hot gas
and their dependence on the host properties. 
In particular, the amount of hot gas in these systems is found to 
span over two orders of magnitude for a given host stellar mass (e.g.,
O'Sullivan, Forbes \& Ponman 2001; 
Ellis \& O'Sullivan 2006), but the origin of this dispersion remains
unclear. It is widely
accepted that the hot gas originates either
externally from accretion of the intergalactic medium (IGM), or internally
from mass loss during normal stellar evolution.
The hot gas content is thus likely dependent on
galaxy age (Sansom et al.~2006), but also depends
sensitively on the formation history and environment of the host
galaxy (Tang et al.~2009a) as well as on-going nuclear activity
(e.g., Lanz et al.~2010).

Galactic bulges are thought to resemble elliptical galaxies in their
stellar populations (i.e., primarily old; e.g., MacArthur, Gonz\'{a}lez \& Courteau 2009), but
a similar assessment has not been made
for their ISM. For instance, it is not clear whether
bulges exhibit the same dispersion in their hot gas content as
in elliptical galaxies. 
This lack of knowledge is due partly to the limited observational information.
Indeed, while the majority of nearby ($D\lesssim$
40 Mpc) optically bright ($m_B$$\lesssim$12) elliptical galaxies have been observed by 
{\sl Chandra} and/or {\sl XMM-Newton}, less than
half of the optically bright early-type spiral galaxies (Sa/Sab; $\sim$60 in total,
Tully 1988) 
in the same volume have
received attention by X-ray pointed observations, and just a few have deep
exposures. Consequently, diffuse X-ray emission has been examined in detail
in only several nearby bulges (M81, Sab, Swartz et al.~2002; 
NGC1291, S0/a, Irwin et al.~2002;
M104, Sa, Li, Wang \& Hameed, 2007, hereafter LWH07; M31, Sb, Li \&
Wang 2007; Bogd\'{a}n \& Gilfanov 2008), for which there
are indications
that the hot gas content is similar to that found in gas-poor
elliptical galaxies (e.g., NGC4697, Sarazin et al.~2001).
However, a definitive view of the properties of hot gas necessarily
awaits deep X-ray observations of an extensive sample of nearby bulges. 

Several facts make the 
Sombrero galaxy (M104=NGC4594; Table \ref{tab:M104}) an excellent target
for an in-depth study of diffuse hot gas in and around a bulge: (i) It
harbors a prominent stellar bulge ($M_B$$\approx$-21.4) as luminous as massive elliptical galaxies 
%massive (with a circular rotation speed of $\sim$$370 {\rm~km~s^{-1}}$)
and thus is expected to host a substantial amount of hot gas; 
(ii) With a well-determined 
distance of $9.0 \pm 0.1$~Mpc, the galaxy allows for 
simultaneously good sensitivity and substantial coverage (out to
$\sim$25 kpc)
that are crucial for tracing extended, low-intensity X-ray emission;
(iii) The nearly edge-on (84$^\circ$) view of
the galaxy allows for a clean separation between its disk and bulge; 
iv) As indicated by the 
low specific fluxes of its diffuse radio, far-infrared and H$\alpha$ emission,
the galaxy shows little indication for recent star formation,
minimizing the possibility of heating and/or gas ejection
from the disk; (v) The galaxy is relatively isolated
and thus uncertainties resulting from galaxy interactions are minimal;
and (vi) The galaxy is located in a direction of relatively
low Galactic foreground absorption column density ($N_{\rm HI} =
3.7 \times 10^{20}$~cm$^{-2}$),
essential for studying the soft X-ray emission from hot gas.
%Therefore, Sombrero is particularly well-suited for
%an X-ray study of high-energy stellar and interstellar products in a
%galactic bulge and their relationship to the galactic disk and to the
%intergalactic environment.

Based on shallow {\sl Chandra} and {\sl XMM-Newton} observations, 
LWH07 reported the detection of 
diffuse X-ray emission from Sombrero, which appears significantly more 
extended than the optical starlight. LWH07 
took advantage of the {\sl Chandra} angular resolution to 
study the inner galactic regions, where many discrete sources are
present and have to be removed,
and of the {\sl XMM-Newton} field of view to trace the diffuse emission
from beyond the galaxy's optical extent. 
Nevertheless, the study of LWH07 was subject to several limitations:
(i) The {\sl XMM-Newton} observation, limited by its relatively large
point-spread function (PSF)
and high instrumental background, is far from ideal for quantifying
the faint, extended X-ray emission; (ii) The instrumental heterogeneity
also challenges the calibration of {\sl Chandra} and {\sl XMM-Newton}
measurements at transition regions;
(iii) Even in {\sl Chandra} images with the bright discrete sources
removed, the apparently ``diffuse'' X-ray
emission is still contaminated by the emission from unresolved stellar objects.
A common practice to account for this stellar component in the
unresolved emission has been to scale the detected hard X-ray photons (e.g.,
with energies above 2
keV), which presumably exclusively arise from stellar objects, to
predict the soft
photons from the same population, by assuming an average hardness ratio
the same as that of the resolved bright sources, primarily
low-mass X-ray binaries (LMXBs).
Not until recently has such an assumption been recognized to be problematic, following
the calibration of the collective X-ray emission from fainter stellar 
populations, i.e.,
coronally active binaries and cataclysmic variables (ABs and CVs;
Sazonov et al.~2006; Revnivtsev et al.~2007, 2008).
These sources exhibit a much softer collective spectrum than that of
LMXBs, such that they can overwhelm faint LMXBs (i.e., those below the
source detection threshold) in terms of contribution to
the unresolved soft X-ray emission. An appropriate treatment is therefore needed to 
decompose the truly diffuse emission, fully accounting for the
collective emission of ABs and CVs.
 
Mindful of the above concerns, we have recently obtained two new {\sl
  Chandra} observations of the Sombrero galaxy
to advance our view of high-energy stellar and interstellar phenomena in a
galactic bulge. Based on the new data, Li et al.~(2010, hereafter Paper I) studied LMXBs in Sombrero. The present work focuses on the
X-ray emission from the diffuse hot gas and from the nucleus.
\S~\ref{sec:data} describes the data preparation toward extracting the extended, unresolved X-ray emission. 
\S~\ref{sec:anal} presents the step-by-step analysis of the nuclear
  emission and the extended emission on various scales, with an emphasis
on the procedure of isolating the truly diffuse emission from the hot gas. 
 \S~\ref{sec:disc} discusses the origin and fate of the hot gas,
  particularly in the scope of a galactic-scale 
outflow.
The study is summarized in \S~\ref{sec:sum}.
Quoted errors are at the 90\%
confidence level throughout this work unless otherwise noted.

\begin{deluxetable}{lr}
\tabletypesize{\footnotesize}
\tablecaption{Basic Information of Sombrero}
\tablewidth{0pt}
\tablehead{
%\colhead{Parameter} &
%\colhead{M~104}
}
\startdata
Morphology$^a$     \dotfill & SA(s)a \\
Center position$^a$ \dotfill & R.A.~$12^{\rm h} 39^{\rm m} 59\fs43$  \\
~~ (J2000) \dotfill & Dec.~$-11^\circ 37^\prime 23\farcs0$ \\
$D_{25}$$^a$\dotfill & $8\farcm7\times3\farcm5$ \\
Inclination angle$^b$  \dotfill & 84$^\circ$\\
%Position angle$^c$ \dotfill & 358$^{\circ}$ \\
B-band magnitude $^a$  \dotfill &  8.98 \\
V-band magnitude $^a$ \dotfill & 8.00 \\
K-band magnitude $^a$ \dotfill & 4.96 \\
%60 ${\mu}$m flux (Jy)$^d$ \dotfill & 7.48 \\
%100  ${\mu}$m flux (Jy)$^d$ \dotfill & 25.86\\ 
Circular speed (${\rm km~s^{-1}}$)$^c$\dotfill  & $370$\\
Distance (Mpc)$^d$ \dotfill & $9.0\pm0.1$ \\
\dotfill & ($1^\prime~\cor~$~2.6kpc) \\
Redshift$^a$ \dotfill & $0.00342$\\
Galactic foreground $N_{\rm \HI}$ ($10^{20}{\rm~cm^{-2}}$)$^e$\dotfill &$3.7$ \\
\enddata
\tablerefs{
 $^a$NED;
 $^b$Rubin et al.~(1985);
 $^c$Bajaja et al.~(1984);
 $^d$Spitler et al.~(2006);
 $^e$Dickey \& Lockman (1990).
}
\label{tab:M104}
\end{deluxetable}

\section{Data preparation} {\label{sec:data}}

{\sl Chandra} observed Sombrero on four occasions over a decade. 
The first observation (ObsID~407; PI: G.~Garmire) was taken on
December 20, 1999, in
a 1/8 sub-array mode with a short exposure of 1.8 ks. The data helped 
demonstrate 
the presence of a bright X-ray source coincident with the galactic nucleus
(Ho et al.~2001; Pellegrini et al.~2002).
The second observation was taken on May 31, 2001 (ObsID~1586; PI:
S.~Murray), with a 19-ks exposure in the ACIS-S FAINT mode, which 
was used to study the X-ray source population (e.g., Di Stefano et
al. 2003) and the hot gas (LWH07).
On April 29 and December 2, 2008 we obtained two new 
observations (ObsID~9532 and 9533; PI: C.~Jones), in the ACIS-I VFAINT mode 
with exposures of 
85 and 90 ks, respectively. In Paper I, data of ObsID 1586, 9532
and 9533 are used to study the discrete X-ray sources. 
In this work we utilize these three observations to study the diffuse emission,
chiefly based on data from the ACIS-I CCDs of
ObsID 9532 and 9533 and from the S3 CCD of ObsID 1586. We also use data 
from the S2 CCD of ObsID 9532 and 9533 to better constrain the local sky background
in our spectral analysis.
In addition, to study the nucleus, we include data from ObsID~407,
taking advantage of its shorter readout frame time (0.4 second,
compared to the standard 3.2 second). 

We reprocessed the data using CIAO v.4.1 and the corresponding calibration 
files, following the {\sl Chandra} ACIS data analysis guide. 
Examination of the light curves indicates that the instrumental
background of each observation was fairly quiescent. Subsequently, 
background filtering is only necessary for ObsID 1586, which results
in 16.7 ks of good time intervals for this data set. 
For each observation, we produced count and exposure
maps in the 0.4-0.7, 0.7-1, 1-2 and 2-6 keV bands, using
a spatial binning factor of 2 ($\sim$$1^{\prime\prime}$ per image 
pixel). 
An absorbed power-law spectrum,
with a photon index of 1.7 and an absorption column density
$N_{\rm H}=10^{21}{\rm~cm^{-2}}$, was adopted as a weighting function when
producing the exposure maps. The energy-dependent difference of 
effective area between
the ACIS-S3 CCD and the ACIS-I CCDs was also accounted for, so that
the quoted count rates throughout this work are referred to ACIS-I. 
Corresponding instrumental background maps were generated from 
the ``stowed'' data, after calibrating with
the 10-12 keV count rate.
These individual maps were then projected to a common
tangential point, here the optical center of Sombrero, to produce summed
images of the combined field of view (FoV; cf.~Fig.~1 of Paper I).  
The total effective exposure, in the 2-6
keV band for instance, is $\gtrsim$180 ks within a projected
galactocentric radius $R \approx 4^\prime$, where the FoV is common to the
three observations (ObsID 1586, 9532 and 9533), and gradually drops below 80 ks at
$R \gtrsim 8^\prime$. 

We detected 383 discrete sources across the combined FoV, the
bulk of which are LMXBs associated with Sombrero, as well as background AGNs
(Paper I). To isolate the unresolved X-ray emission,
we have excluded these sources from maps in individual observations,
each with a circle enclosing $\sim$$96\%$ of the source photons.
Such a source subtraction procedure works well except for the bright nucleus,
whose PSF-scattered photons contribute substantially to the
unresolved emission, especially at energies above 2 keV, within the central half-arcminute. We therefore relied on
{\sl MARX} simulations (see \S~\ref{subsec:nuc}) to determine the
expected distributions of PSF-scattered photons from the nucleus.
Residual PSF-scattered photons from other sources are estimated to
contribute $\sim$25\% of the remaining unresolved X-ray emission in
the 2-6 keV band.

Due to the variation of the local background, PSF and effective
exposure across the FoV, the limiting flux for source detection 
necessarily varies with sky position, 
roughly increasing with angular distance from the galactic center.  
Consequently, at larger radii the background contains more photons from unresolved sources.
We statistically corrected for such a non-uniform background, based on 
the luminosity functions of the LMXBs and cosmic AGNs (Paper I), to a common
source flux limit of $10^{-15}{\rm~ergs~s^{-1}~cm^{-2}}$ (0.4-6 keV)
across the FoV.

We also accounted for artifacts due to photons
registered during CCD readout, following the technique
proposed by Markevitch et al.~(2000). The procedure effectively generated
maps of statistically-sampled readout photons for individual
observations, which can then be projected in the same way as for the
count maps.

The three terms considered above, namely, the PSF wing of the nucleus,
the spatially non-uniform 
photons from unresolved sources, and the out-of-time events, all contribute to the unresolved
 emission depending on the energy band and sky position.
We added these three terms to the instrumental background to
form a fiducial background to be subtracted. A presumably uniform
sky background, e.g., from Galactic foreground emission,
is then determined locally (see \S~\ref{subsec:spec}).

\section{Analysis and results} {\label{sec:anal}}

\subsection{The X-ray nucleus} {\label{subsec:nuc}}

\begin{figure*}[!htb]
\centerline{
\epsfig{figure=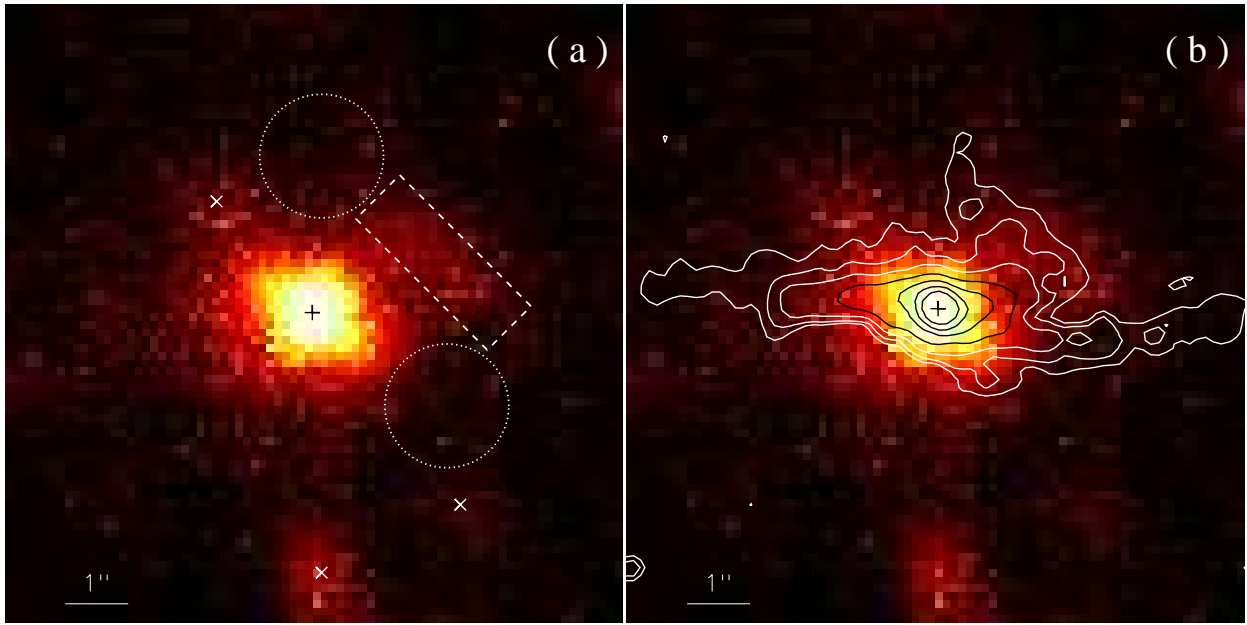,width=\textwidth,angle=0,clip=}
}
\caption{A 0.3-8 keV count image from the nucleus and its
  $10^{\prime\prime}\times10^{\prime\prime}$ ($\sim$430 pc by 430 pc)
  vicinity, on a pixel scale of one fourth of the
  ACIS pixel size 0\farcs492.
  In (a), the  $3^{\prime\prime}\times1^{\prime\prime}$ dashed rectangle
  outlines the region where the spectra of an apparently extended
  X-ray feature are extracted, while the two dotted circles show the
  adopted background region.
  %{\sl Right}: A $V-I$ color map obtained
  %from {\sl HST} observations, showing the circumnuclear dust
  %lanes. The VLA 8.4 GHz continuum emission is also shown in
  %contours. 
  The centroid of the nucleus is marked by a plus sign, while crosses highlight
  three point-like X-ray sources. In (b), intensity contours of the
  H$\alpha$+[N II] emission obtained from {\sl HST} observations are
  plotted. Note the positional coincidence between the extended X-ray
  feature and the optical line emission northwest of the nucleus. 
}
\label{fig:nuc_mul}
\end{figure*}

\begin{figure*}[!htb]
\centerline{
\epsfig{figure=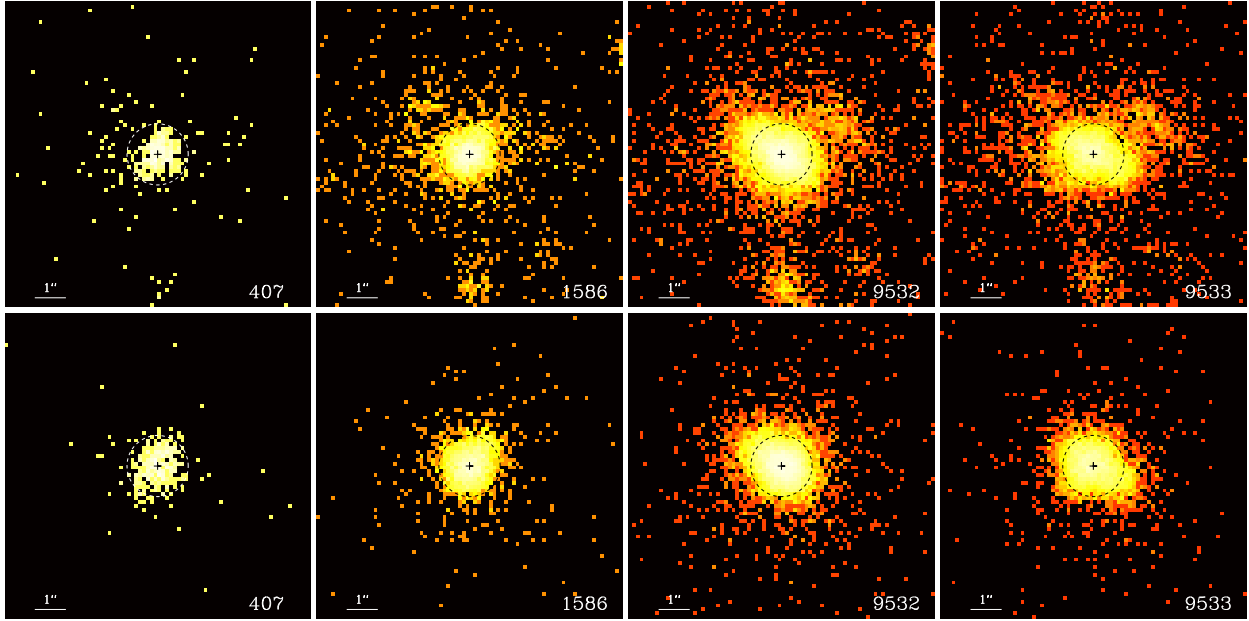,width=\textwidth,angle=0,clip=}
}
%\centerline{
%\epsfig{figure=nucleus_407_sim.ps,width=0.25\textwidth,angle=90,clip=} 
%\epsfig{figure=nucleus_1586_sim.ps,width=0.25\textwidth,angle=90,clip=} 
%\epsfig{figure=nucleus_9532_sim.ps,width=0.25\textwidth,angle=90,clip=} 
%\epsfig{figure=nucleus_9533_sim.ps,width=0.25\textwidth,angle=90,clip=} 
%}
\caption{{\sl Upper panel}: 0.3-8 keV X-ray emission from the nucleus and its
  $10^{\prime\prime}\times10^{\prime\prime}$ ($\sim$430 pc by 430 pc) vicinity, detected in four
{\sl Chandra} observations, on a pixel scale of one fourth of the
  ACIS pixel size 0\farcs492. The dashed circle outlines the region
  where the spectra of the nucleus are extracted.
  {\sl Lower panel}: Simulated 0.3-8 keV X-ray emission from
  the nucleus for the four observations, using the CIAO tool {\sl ChaRT}.
  The centroid of the nucleus is marked by a plus sign.}
\label{fig:nuc}
\end{figure*}

Sombrero hosts a super-massive black hole
(SMBH) with a mass of                         
$\sim$10$^{9}{\rm~M_\odot}$ (Kormendy
1988; Kormendy et al.~1996) inferred from its central
stellar kinematics.
Classified as a LINER (low-ionization nuclear emission region; Heckman 1980; Ho, Filippenko, Sargent 1997), Sombrero
is known to exhibit a low level of nuclear activity. In particular, 
the nucleus (i.e., the SMBH) manifests itself as a moderately bright X-ray source
(Fabbiano \&
Juda 1997; Ho et al.~2001), as clearly seen in Fig.~\ref{fig:nuc_mul}, a stacked image (i.e., combining
the four {\sl Chandra} observations) 
of the 0.3-8 keV emission from the central 10$^{\prime\prime}$ by 10$^{\prime\prime}$ 
region of Sombrero.
The upper panel of Fig.~\ref{fig:nuc} further shows the same region as 
detected in the individual observations. 
These images are shown on a refined pixel scale (1/4 of the original
ACIS pixel) to take advantage of the sub-pixel positioning information
that results from telescope dithering.
%Clearly, a bright source coincident with the nucleus dominates the emission. 
The sub-arcsec resolution afforded by {\sl Chandra} is ideal for disentangling 
the nucleus. Nevertheless, quantification of
the nuclear emission demands caution on the effects of photon pile-up and PSF
scattering.

The moderately bright nucleus (with a flux of
$\sim$$10^{-12}{\rm~ergs~s^{-1}}$) is expected to cause mild pile-up.
Direct evidence of pile-up
comes from examination of events flagged as grade 7 in the
raw (i.e., level 1) event file (the reprocessed, i.e., level
2, data keeps only ``good'' events with grades 0, 2, 3, 4, 6). When pile-up occurs,
i.e., two or more photons land on the same CCD frame and consequently
only one event is registered, a grade-7 event can be produced due to,
for instance,
a combination of grades 2-2 or grades 3-4. Visual inspection reveals that
grade-7 events are clustered in the central 3 pixels around the
nucleus in ObsID 1586, 9532 and 9533, but are essentially absent in
ObsID 407, presumably due to the much shorter frame time of
the latter.

\begin{figure*}[!htb]
\centerline{
\epsfig{figure=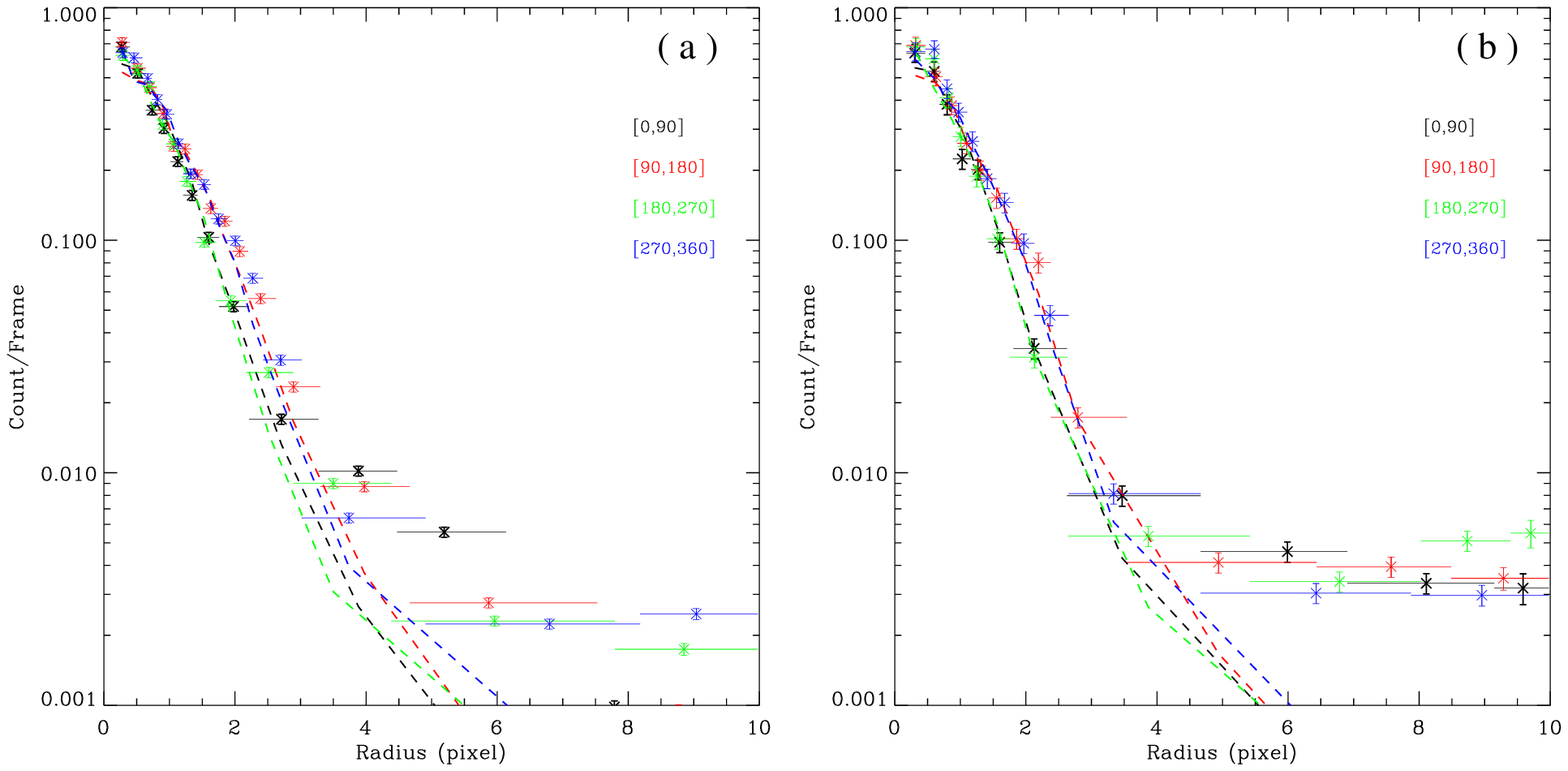,width=\textwidth,angle=0,clip=}
}
\caption{(a) 0.3-8 keV radial intensity profiles around the
  nucleus (color-coded data points, adaptively binned to achieve 
  a minimum of 400 counts per bin), extracted from the combined image
  of the four observations, for four quadrants
  whose angular ranges (measured counterclockwise from west) are
  color-coded as shown. The curves are the corresponding profiles extracted from
the {\sl ChaRT}-simulated image. The vertical axis is in units of 
  count/frame, where one frame is defined as 3 pixel$\times$3
  pixel$\times$3.2 s.
(b) Similar to (a), but the data
  points, adaptively binned to achieve a minimum of
  100 counts per bin, are from
  stacking a few bright discrete sources located in the
  vicinity of the nucleus and then multiplied by a factor of 5 to match the
  intensity profile of the nucleus. Vertical error bars are of 1$\sigma$.}
\label{fig:nuc_rsb}
\end{figure*}

We examine the PSF by using
simulations provided by the CIAO tools {\sl ChaRT}\footnote{http://cxc.harvard.edu/chart/} and {\sl MARX}\footnote{http://space.mit.edu/CXC/MARX/}. By taking the
incident source
spectrum (based on {\sl a posteriori} knowledge from
the spectral analysis; see below), the source location on the detector and the
desired exposure, {\sl ChaRT} traces simulated rays that can
be processed by {\sl MARX} to generate simulated events, as if
registered in a real observation. The effect of pile-up can also be
mimicked by {\sl MARX}, based on the algorithm of Davis (2001).
0.3-8 keV events from the nucleus are thus simulated for the four individual observations and are shown in the
lower panel of Fig.~\ref{fig:nuc}. General agreement between the
observed and simulated images is readily seen, especially in
ObsID~9532 and 9533 where the nucleus is located at a relatively
large off-axis angle ($\sim$2$^\prime$) and hence the PSF elongation
along a northeast-southwest direction is significant.
More quantitatively, we compare the observed and
simulated radial intensity profiles, shown in Fig.~\ref{fig:nuc_rsb}a,
which are constructed by combining the 0.3-8 keV
counts registered in individual observations for each of the
four quadrants. The second (northeast;
90$^\circ$-180$^\circ$) and fourth (southwest; 270$^\circ$-360$^\circ$) quadrants
show higher intensities, again due to the PSF elongation. The
simulated profiles follow the observed profiles in all four quadrants
in the central two pixels ($\sim$1$^{\prime\prime}$) where $\sim$80\% of
the nuclear flux is enclosed; deviations become substantial at larger
off-center distances presumably due to the presence of extra-nuclear
emission (e.g., arising from discrete sources and diffuse hot gas; see
\S~\ref{subsec:circum}) that is not simulated.
The goodness of the simulated PSF is further supported in
Fig.~\ref{fig:nuc_rsb}b, where it is compared with the
intensity profile obtained by stacking several
discrete sources located within $\sim$20$^{\prime\prime}$ of the
nucleus. The stacked intensity profile is expected to represent the actual
PSF at the position of the nucleus, averaged among the four
observations. Taking into account the local background, the
stacked and simulated profiles agree well with each other, except
in the central pixel, presumably due to the fact that the discrete
sources are individually free of photon pile-up, while the much brighter
nucleus is mildly so.

Next we quantify the spectral property of the nucleus.
Spectra are extracted from a 2-pixel (1$^{\prime\prime}$) radius
circle around the nucleus for individual
observations. Fig.~\ref{fig:nuc_rsb} indicates that the local
background can be neglected for this relatively small aperture.
An absorbed power-law (PL), modified by the pile-up
model of Davis (2001), is used to fit the spectra. The
spectra were first fitted individually. We found that an acceptable
fit was obtained for each spectrum and the resultant photon indice
($\Gamma$) and absorption column densities ($N_{\rm H}$) are consistent
with each other within statistical uncertainties. We note that the inclusion
of the pile-up model leads to a steeper PL (i.e., a larger
photon index) for the spectra of ObsID 1586, 9532 and 9533, but not
for the spectrum of ObsID 407. This supports the above argument that
the nucleus is not subject to pile-up in ObsID 407.
We then fit the spectra
simultaneously, forcing both $\Gamma$ and $N_{\rm H}$ to be identical
in order to reduce the statistical uncertainty. The fit was acceptable with a
$\chi^2$/dof = 526.7/474, giving $\Gamma =
1.98^{+0.08}_{-0.12}$ and $N_{\rm H} =
3.4^{+0.1}_{-0.2}\times10^{21}{\rm~cm^{-2}}$.
Consistent fit results on these two parameters have been reported,
based on, e.g., the {\sl XMM-Newton} (Pellegrini et al.~2003), {\sl BeppoSAX}
(Pellegrini et al.~2002) and {\sl ASCA} (Nicholson et al.~1998)
spectra that are of much poorer angular resolutions and necessarily
contaminated by extra-nuclear emission that requires additional spectral
components.
%The four ACIS spectra and the best-fit models are shown in
%Fig.~\ref{fig:nuc_spec}.

%\begin{figure*}[!htb]
%\centerline{
%\epsfig{figure=nuc_spec.ps,width=0.6\textwidth,angle=270,clip=}
%}
%\caption{0.3-8 keV spectra of the nucleus extracted from individual
%    observations. {\sl Blue}: ObsId.~407; {\sl Black}: ObsID.~1586;
%    {\sl Red}: ObsID.~9532; {\sl Green}: ObsID.~9533. The spectra are adaptively
%    binned to achieve a signal-to-noise radio greater than 4 and a
%    minimum of 30 counts per bin. Also
%    shown is the best-fit absorbed power-law model subject to
%    pileup. See text for details.}
%\label{fig:nuc_spec}
%\end{figure*}

In the fit the normalization was allowed to vary among the four spectra,
resulting in 3.9$^{+0.4}_{-0.4}$, 3.7$^{+0.3}_{-0.3}$, 3.8$^{+0.1}_{-0.1}$ and 1.9$^{+0.1}_{-0.2}\times10^{-4}$ for ObsID 407, 1586,
9532 and 9533, respectively. Therefore during the last observation, ObsID
9533, the nucleus shows a factor of two lower flux than in the first
three observations. Corrected for absorption and the enclosed energy fraction within
the spectral extraction region, which varies between 76\%-83\%
according to the simulated PSFs for the four observations, the 0.3-8 keV
intrinsic nuclear flux is $\sim$2.5, 2.4, 2.6 and 1.3
$\times10^{-12}{\rm~ergs~s^{-1}~cm^{-2}}$ in ObsID 407, 1586, 9532 and
9533, respectively. This means that the X-ray luminosity of the nucleus might have been 
constant at $\sim$$2.4\times10^{40}{\rm~ergs~s^{-1}}$
between December 1999 and April 2008, but then dropped by a factor of two in 7 months.
%An X-ray light curve of the nucleus is presented in Fig.~\ref{fig:lc}, based
%on measurements from the literature and this work. 
%An energy range of 2-10 keV is chosen to reduce the contamination arising from
%extra-nuclear emission inevitably enclosed in the aperture of non-{\sl
%  Chandra} detectors. The corresponding flux values are derived based on the
%best-fit power-law models given in the literature.
%We note that even in the hard X-ray band the extra-nuclear emission,
%in particular that arising from low-mass X-ray binaries (Paper I),
%can have a substantial contribution. We estimate an integrated 2-10
%keV luminosity of $\sim$1.4$\times10^{40}{\rm~ergs~s^{-1}}$ for the total
%extra-nuclear emission within a radius of 4 arcmin (roughly the
%optical extent of Sombrero). Therefore, there is no strong evidence
%that the X-ray nucleus showed significant variability, except in 2008.
We also searched for, but found no evidence of, intra-observation
variability of the nucleus in any individual observation.

We note that adding
a Gaussian line at 6.4 keV or 6.7 keV (i.e., from iron K$\alpha$) does not statistically
improve the best-fit absorbed PL model. We constrain
the equivalent width (EW) of a possible Fe K$\alpha$ line by setting the normalization
of the Gaussian such that the $\chi^2$ is increased by a value of
9. For a 6.4 keV line, we obtain a 3$\sigma$ upper limit on the EW of 110 eV for a narrow
line (assuming $\sigma_E$=0) and of 190 eV for a broad line
(assuming $\sigma_E$=0.5 keV). For a 6.7 keV line, the corresponding values are
240 eV (narrow line) and 280 eV (broad line).

%\begin{figure*}[!htb]
%\centerline{
%\epsfig{figure=nucleus_lc.ps,width=0.6\textwidth,angle=90,clip=}
%}
%\caption{A 2-10 keV light curve of the nucleus. The {\sl Chandra}
%  measurements are derived from this work. Other measurements are collected
%  from the literature, rescaled to the adopted distance and converted
%  to the considered energy range, when necessary.}
%\label{fig:lc}
%\end{figure*}

\subsection{The circumnuclear X-ray emission} {\label{subsec:circum}}
%The 0.3-8 keV mosaicked image (i.e., combining the four observations) of the
%nuclear region is shown in Fig.~\ref{fig:nuc_mul}a. 
Several features are clearly present in the circumnuclear region,
including three discrete sources (each marked by a cross in
Fig.~\ref{fig:nuc_mul}a) and an apparently extended feature northwest
of the nucleus  (approximately
enclosed by a dashed rectangle in
Fig.~\ref{fig:nuc_mul}a). 
%The latter is particularly interesting owing
%to its position relative to the nucleus, which is closely alligned with the radio jet. 
We extract spectra for the extended feature from all observations except ObsID 407, from which 
we can only estimate $4.7\pm3.2$ (1$\sigma$) net counts, or a net count rate of 
$(2.6\pm1.8)\times10^{-3}{\rm~cts~s^{-1}}$  for the feature.
We note that the adopted background region (defined by two
dotted circles in Fig.~\ref{fig:nuc_mul}) encloses approximately the same
amount of the PSF-scattered nuclear flux per unit area as in the
rectangle. The spectra were first fit individually with an
absorbed PL model. The resultant parameters are statistically
consistent among the three spectra. Hence we fit the spectra
simultaneously, linking all parameters. The fit is acceptable with a
$\chi^2$/dof = 33.4/30, giving $\Gamma =
1.64^{+0.26}_{-0.18}$, $N_{\rm H} =
1.4^{+1.4}_{-0.7}\times10^{21}{\rm~cm^{-2}}$, and a 0.3-8 keV
intrinsic luminosity of $3.6\times10^{38}{\rm~ergs~s^{-1}}$.
This model predicts a net count rate of $6.2
\times10^{-3}{\rm~cts~s^{-1}}$ to be measured in ObsID 407, i.e., $\sim$2$\sigma$
higher than the observed value.

The extended X-ray feature is particularly interesting in view of its possible relation
with the nucleus. Gallimore et al.~(2006) found  
in Sombrero a kpc-scale linear radio structure,
which they suggested originates from a SMBH-powered jet/outflow.
The position of the X-ray feature relative to the nucleus is
coincident with that of the radio structure, i.e., at a position 
angle $\sim$45$^\circ$ west of north.
We investigated the Very Long Array archival data on arcsec-scales
but found no evidence for extended radio emission coincident with the X-ray feature.
Alternatively, the X-ray feature
may originate from scattered nuclear emission and/or
a photoionized gas, which requires the
presence of circumnuclear dusty gas at the location of the X-ray feature. 
{\sl HST} observations have resolved the narrow line region (NLR) in
H$\alpha$+[N II] emission, which mainly consists of a core and two arms lying
east-west, i.e., in the edge-on disk 
(Emsellem \& Ferruit 2000;
Masegosa et al.~2010; Fig.~\ref{fig:nuc_mul}b). 
Another curved H$\alpha$+[N II] feature is evident northwest of the nucleus, and
interestingly, is roughly coincident with the extended X-ray feature
(Fig.~\ref{fig:nuc_mul}b). The two arms, but not the curved feature,
are also clearly seen as dust extinction against the starlight
(Emsellem \& Ferruit 2000). If both the H$\alpha$+[N II] and X-ray
emission arise from a photoionized gas, the extended feature is likely
associated with the hypothesized jet/outflow and is thus located out
of the disk. However, this is not readily evident in the NLR
kinematics (Emsellem \& Ferruit 2000). 
For reference, we estimate an H$\alpha$+[N II] luminosity of
$\sim$$5\times10^{38}{\rm~erg~s^{-1}}$ for the feature, which is
comparable to the X-ray luminosity from the same region. When associated X-ray emission is
detected, NLRs often exhibit comparable luminosities in
their X-ray and optical line emission (e.g., Bianchi, Guainazzi \&
Chiaberge 2006). Nevertheless, we note that the spectra of the X-ray
feature show no sign of emission lines that are expected to arise from
a photoionized gas.  On the other hand,
if powered by star formation, the H$\alpha$+[N II]
luminosity indicates a star formation rate of only
$\sim$$10^{-4}{\rm~M_\odot~yr^{-1}}$ (e.g, Kennicutt 1998), 
which is too low to produce the observed X-ray emission from the
feature.
Lastly, the X-ray feature may be overlapping stellar sources.
This is consistent with the fitted photon index of $\sim$1.6,
typical of LMXBs, but requires at least two LMXBs of 
near-Eddington luminosity (for an accreting neutron star) coincidentally
aligned. The lower count rate measured in ObsID
407 would further imply flux variability for both sources.
Future radio/optical/X-ray observations may help distinguish the above
scenarios for this interesting feature.

To characterize the properties of the circumnuclear hot gas, we further extracted 
spectra from a 3$^{\prime\prime}$-5$^{\prime\prime}$ annulus around
the nucleus, from ObsID 1586, 9532 and 9533, in which discrete sources
and the extended feature are excluded. For simplicity, we use the
FTOOL {\sl addspec} to coadd the spectra. The
coadded spectrum is fit by an absorbed MEKAL+PL model. The PL component, with a fixed photon index of 1.9, is applied to represent the contributions 
from the PSF-scatted nuclear flux and the residual flux of stellar sources. 
The MEKAL component, with the abundance fixed at solar, constrains the diffuse hot gas. 
The fit is acceptable with $\chi^2$/dof = 24.6/19, giving a gas temperature
of 0.52$^{+0.12}_{-0.22}$ keV and a gas density of 0.13$^{+0.04}_{-0.02}{\rm~cm^{-3}}$,
assuming that the diffuse emission comes from gas filling a spherical shell. 
Our results are consistent with the finding of Pellegrini et al.~(2003)
based on the data of ObsID 1586. The gas has a cooling timescale of $\sim$$5\times10^7$ yr, 
according to the cooling function of Sutherland \& Dopita (1993). 

Assuming that the nucleus is powered
by Bondi (1952) accretion of the circumnuclear hot gas,
the accretion rate can be estimated as
\begin{eqnarray}
\dot{M}_{\rm Bondi} \approx 4{\pi}{\lambda}{\mu}m_{\rm H}n(GM_{\rm BH})^2c_s^{-3}~~~~~~~~~~~~~~~~~~~~~~~~~~~~~~~~~~~~~~~~~~~~ \\ \nonumber
\approx 1.8\times10^{-3}(\frac{n}{0.1{\rm~cm^{-3}}})(\frac{M_{\rm BH}}{10^9{\rm~M_\odot}})^2(\frac{kT}{0.5{\rm~keV}})^{-1.5}{\rm~M_\odot~yr^{-1}},
\end{eqnarray}
where ${\lambda}$ is a numerical factor taken to be 0.25, $c_s = ({\gamma}kT/{\mu}m_{\rm H})^{1/2}$ is the sound speed, and the remaining symbols have conventional meanings.
The corresponding {\sl Bondi luminosity} is $L_{\rm Bondi} \equiv \eta\dot{M}_{\rm Bondi}c^2 \approx 1.0\times10^{43} (\eta/0.1){\rm~ergs~s^{-1}}$,
where $\eta$ is the radiation efficiency. 
It follows that $L_{\rm X}/L_{\rm Bondi}\sim10^{-3}$, or $L_{\rm X}/L_{\rm Eddington}\sim10^{-7}$,
indicating that the nucleus is radiatively inefficient, as first noted by Pellegrini et al.~(2003). We also note that the {\sl Bondi radius}, $R_B\equiv2GM_{\rm BH}/c_s^{2}\approx105 {\rm~pc}$, spans an angle of 2\farcs4. However, the PSF wing of the bright nucleus hampers the possibility of resolving the accretion flow. 

\subsection{The galactic-scale X-ray emission} {\label{subsec:unr}}
\subsubsection{Broad band distributions} 
\begin{figure*}[!htb]
\centerline{
\epsfig{figure=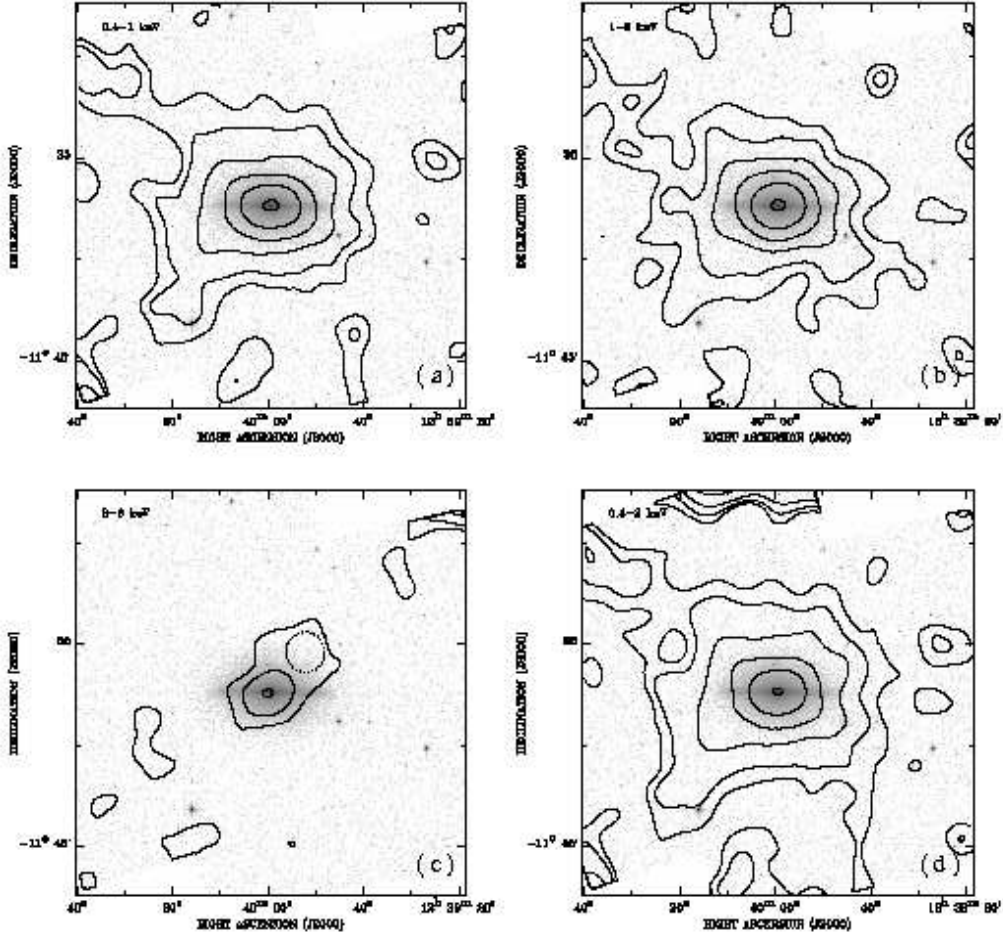,width=\textwidth,angle=0,clip=}
}
%\centerline{
%\epsfig{figure=M104_unr_2-6_K.ps,width=0.5\textwidth,angle=90,clip=} 
%\epsfig{figure=M104_dif_0.4-2_K.ps,width=0.5\textwidth,angle=90,clip=}
%}
\caption{(a)-(c): 2MASS K-band image of Sombrero, overlaid by
   contours of the background-subtracted, exposure-corrected
  unresolved X-ray emission in the 0.4-1, 1-2 and 2-6 keV bands. 
The X-ray emission is smoothed
   by a Gaussian kernel of $\sigma$=40$^{\prime\prime}$ to bring up the faint extended
   emission. The dotted circle in (c) outlines the
   region where anomalous 2-6 keV signals are present. (d): the 0.4-2
   keV diffuse X-ray emission, i.e., the contribution from unresolved
   stellar populations has been subtracted, overlaid on the 2MASS
   K-band image. The contour levels decrease by a factor of 2 in each
   step outward.
} 
\label{fig:unr}
\end{figure*}

{\label{subsubsec:morpho}}
At the galactic scale, the unresolved (i.e., source- and background-subtracted) X-ray
emission in the four bands is shown in Fig.~\ref{fig:unr}, in
comparison with a 2MASS K-band image (Jarrett et al.~2003) that
traces the old stellar populations in Sombrero. 
The 2-6 keV (hereafter hard band) emission
traces the starlight (Fig.~\ref{fig:unr}c), indicating a stellar origin.
Indeed, the contribution of diffuse gas with a typical sub-keV temperature is expected to be 
negligible at energies above 2 keV. 
The hard band emission shows some 
asymmetry toward the northwest (outlined by a circle in Fig.~\ref{fig:unr}d), although we stress caution
when examining a smoothed image. 
%A close examination of the region of asymmetry
%(outlined by a circle in Fig.~\ref{fig:unr}d) reveals that the 2-6 keV
%net count rate varies over a factor of two among the three
%observations. Moreover, in both ObsID 9532 and 9533, the spectral
%shape above 2 keV appears unphysically flat.
We suggest that this asymmetry is due to residual particle contamination or a background object. 

\begin{figure*}[!htb]
\centerline{
\epsfig{figure=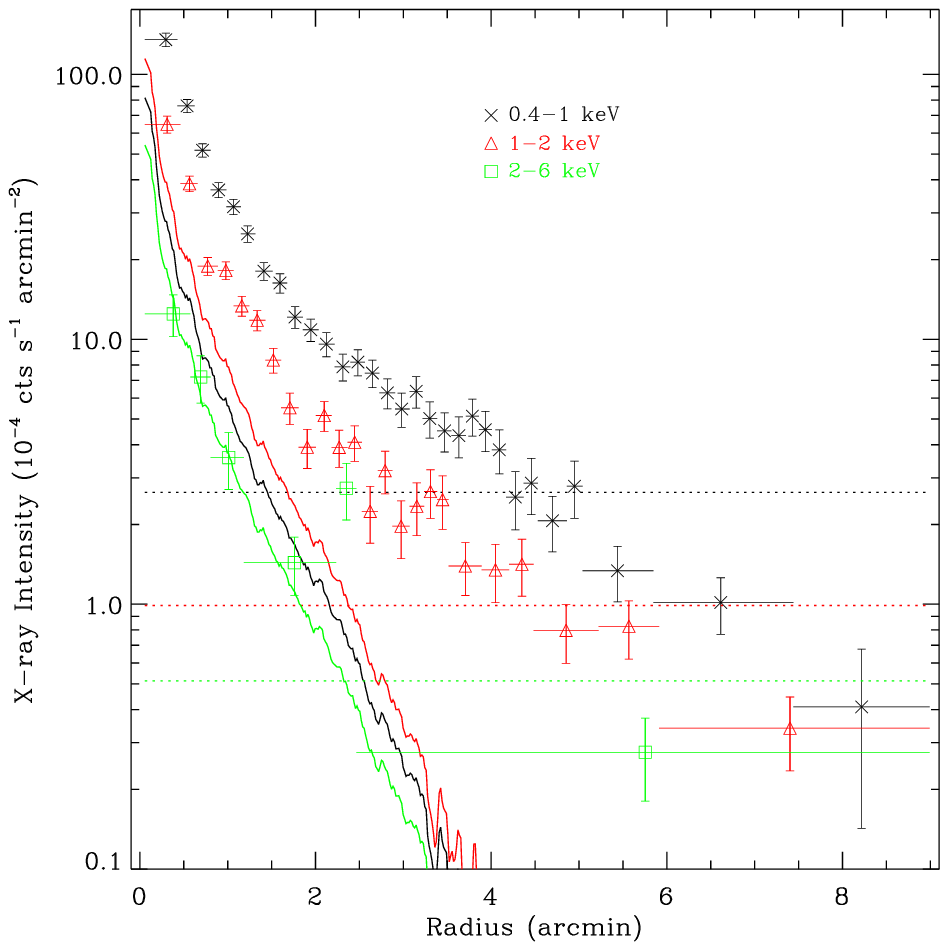,width=0.6\textwidth,angle=0,clip=} 
}
\caption{Azimuthally-averaged radial intensity distributions of the unresolved X-ray emission
  from Sombrero, which are adaptively binned to achieve a
  signal-to-noise (with respect to the subtracted instrumental and sky background) 
  ratio better than 4 and a minimum of 400 counts in each bin.
  Solid curves are normalized K-band intensity profiles, 
  representing the
  combined contribution to individual bands from unresolved LMXBs,
  CVs and ABs. The dotted lines represent the level of the local sky
  background, effectively determined from regions beyond $R=9^\prime$. 
  See text for details.}
\label{fig:rsb}
\end{figure*}

%\begin{figure*}[!htb]
%\centerline{
%\epsfig{figure=blob_spec.ps,width=0.6\textwidth,angle=270,clip=} 
%}
%\caption{Spectra of the extended hard X-ray emission. {\sl Black}:
%  ObsID 1586; {\sl Red}: ObsID 9532; {\sl Green}: ObsID 9533. The
%  spectra are jointly fitted by a {\sl wabs(MEKAL+PL)} model. The best-fit 
%  gives $T = 0.56^{+0.07}_{-0.09}$ and $\gamma = 0.66^{+0.50}_{-0.13}$.}
%\label{fig:blob}
%\end{figure*}
At energies below 2 keV, the X-ray emission 
appears significantly more extended than that of the K-band starlight
(Fig.~\ref{fig:unr}a-b). This
indicates that diffuse hot
gas is present in and around the bulge of Sombrero.
More quantitatively, we compare the radial distributions of the X-ray
emission and the K-band light. As shown in Fig.~\ref{fig:rsb}, 
the hard band radial distribution closely follows that of the K-band light
within $R\approx2^\prime$. Indeed, a normalized K-band profile 
is found to be an adequate characterization of the hard
band distribution (green curve in Fig.~\ref{fig:rsb}; the excess point at $\sim$2\farcm5 is due to the anomalous feature mentioned above). The normalization factor in this case is
$(7.0\pm1.2)\times10^{-5}{\rm~cts~s^{-1}/(MJy~sr^{-1})}$. 
This supports the expectation that the hard band emission arises from old stellar
populations, namely, unresolved LMXBs, CVs and ABs (\S~\ref{sec:intro}).
In particular, the CV+AB populations appear to show a quasi-universal X-ray
emissivity (per unit stellar mass) in the hard band, which has
been calibrated to $\sim$30\% uncertainty (Revnivtsev et
al.~2007). In Sombrero, such a component is expected to account for
$(3.4\pm1.0)\times10^{-5}{\rm~cts~s^{-1}/(MJy~sr^{-1})}$ in the hard
band. The rest, about half, of the hard band emission can be
attributed to unresolved LMXBs and the PSF residuals of resolved LMXBs. 

The stellar contribution to the soft bands can be determined by
scaling the hard band emissivity, with certain assumptions on the integrated
spectral shape of the individual stellar populations. For the LMXB component, an absorbed PL
model, with a photon index of 1.7 and an absorption column density of
$10^{21}{\rm~cm^{-2}}$, is adopted. For the CV+AB component, a fiducial spectral
model based on the extended emission from M32 (Revnivtsev et
al.~2007) is adopted. Specifically, we choose 
a two-temperature thermal plasma model (MEKAL+MEKAL in XSPEC), with
temperatures of 0.38 and 4.6 keV and solar abundances. We note that
this model is somewhat different from the MEKAL+PL model 
adopted by Revnivtsev et al.~(2007), but preserves the idea that CVs and
ABs typically exhibit soft and hard thermal spectra,
respectively. We also note that the universality of the CV+AB
emissivity is less certain at energies below 2 keV. In fact,
M32 shows the lowest measured value of soft X-ray emissivity. 
Other measurements, such as those for NGC3379 and the M31 disk (Revnivtsev et
al.~2008), yield a factor of 2 higher values, although the
presence of soft X-ray-emitting gas cannot be
completely ruled out in those two galaxies.

The stellar contribution in individual bands is thus determined and
shown as curves in Fig.~\ref{fig:rsb}. It is immediately clear
that stellar objects cannot fully account for the observed soft
X-ray emission: {\sl the expected 
stellar emission shows both a lower intensity and a smaller extent
than the observed unresolved emission}. 
We therefore conclude the presence of truly diffuse emission arising from
hot gas in and around the Sombrero bulge.

\subsubsection{Truly diffuse X-ray emission} 
\begin{figure*}[!htb]
\centerline{
\epsfig{figure=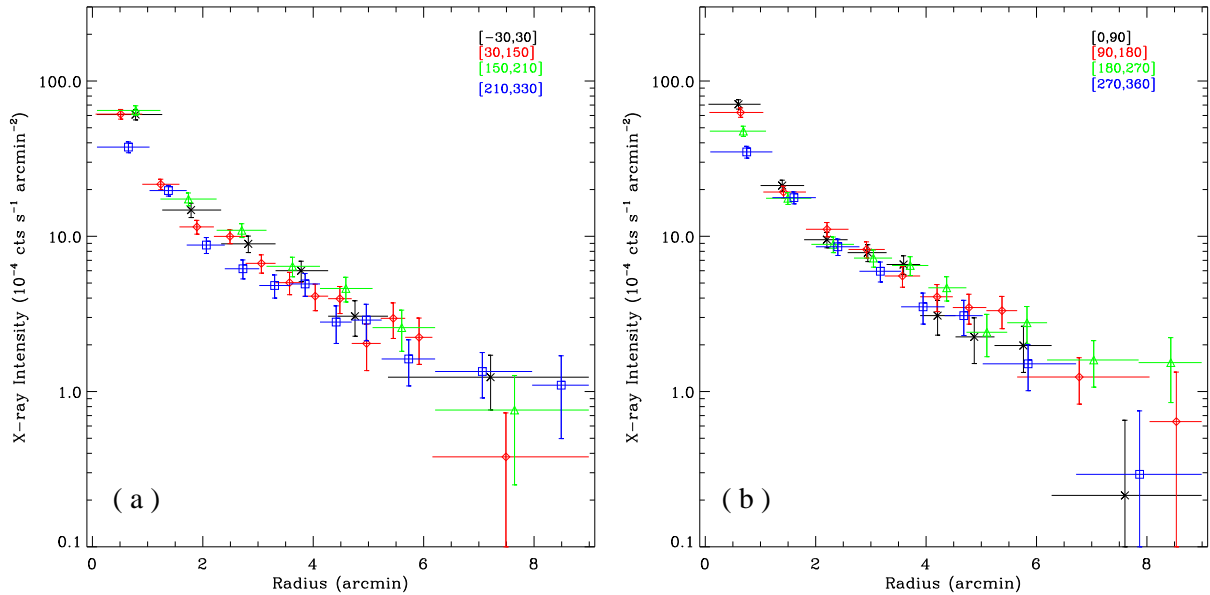,width=\textwidth,angle=0,clip=}
}
\caption{Azimuthally-averaged
radial intensity profiles of the 0.4-2 keV diffuse emission,
constructed from four quadrants and shown in different colors. The
azimuthal ranges of the four quadrants are indicated in each panel.
Angles are measured counterclockwise from the
west, a direction assumed to be aligned with the major-axis of the
disk. The profiles are binned to achieve a signal-to-noise (with
respect to the combined stellar and background emission) better than
3 and a minimum of 900 counts per bin.
}
\label{fig:rsb_dif_4q}
\end{figure*}

{\label{subsubsec:diffuse}}
The 0.4-2 keV diffuse X-ray emission can then be obtained by subtracting the stellar
component from the total unresolved emission (Fig.~\ref{fig:unr}d). 
We construct radial intensity profiles for the diffuse emission in
four quadrants of selected azimuthal ranges (Fig.~\ref{fig:rsb_dif_4q}), chiefly to examine deviations from symmetry. In Fig.~\ref{fig:rsb_dif_4q}a, the quadrants are
defined to approximately follow the major- and minor-axes. In the
central $\sim$4$^\prime$, the diffuse emission appears stronger along
the major-axis (i.e., azimuthal ranges of [-30$^\circ$, 30$^\circ$] and [150$^\circ$,
  180$^\circ$]) than along the minor-axis. In the innermost region,
the emission appears faintest along the southern minor-axis (i.e.,
azimuthal ranges of [210$^\circ$, 330$^\circ$]), consistent with
absorption by the dusty disk, the near side of which is tilted south.
In Fig.~\ref{fig:rsb_dif_4q}b, the quadrants are defined as enclosed
by a pair of the major- and minor-axes. In this configuration, 
the diffuse emission appears rather symmetric within the central
$\sim$4$^\prime$, 
except it is fainter in the southern half of the innermost region,
again presumably due to the absorption by the slightly tilted disk. At large radii (beyond
$\sim$6$^\prime$), the emission appears stronger in the
southeast, as suggested in Fig.~\ref{fig:unr}.

\begin{figure*}[!htb]
\centerline{
\epsfig{figure=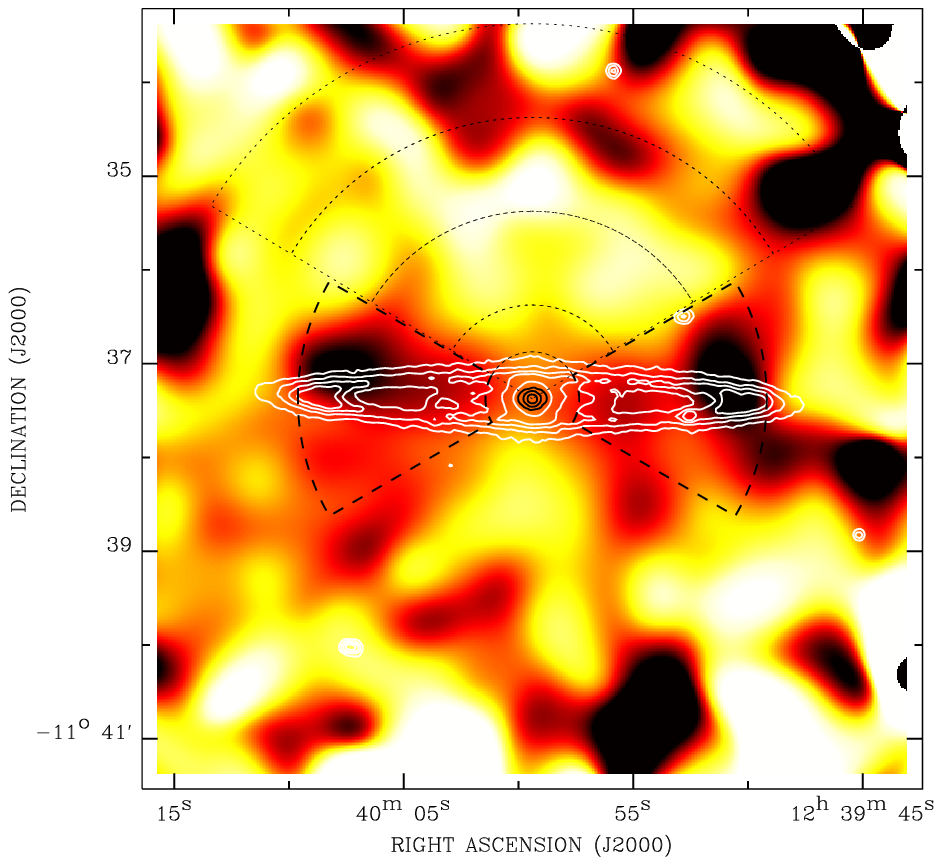,width=0.6\textwidth,angle=0,clip=}
}
%\centerline{
%\epsfig{figure=M104_dif_1-2_K.ps,width=0.5\textwidth,angle=90,clip=} 
%\epsfig{figure=M104_dif_2-6_K.ps,width=0.5\textwidth,angle=90,clip=} 
%}
\caption{
%(a) Gaussian-smoothed contours of the 0.4-2 keV observed diffuse X-ray emission, i.e., the contribution from   unresolved stellar populations has been subtracted, overlaid on the 2MASS K-band image. 
An image of the hardness of the diffuse emission, defined as $(I_{\rm
  1-2~keV}-I_{\rm 0.4-1~keV})/(I_{\rm1-2~keV}+I_{\rm
  0.4-1~keV})$. Darker colors indicate lower values of hardness and
  hence softer emission. The contours show the {\sl Spitzer}/MIPS 
  24 $\mu$m emission that traces the dust lane in the
  disk and the bright nucleus. The two dashed sectors define
  the disk region for spectral analysis, whereas the consecutive dotted sectors
  define inner bulge regions for spectral deprojection (see \S~\ref{subsec:spec}).    
}
\label{fig:dif}
\end{figure*}

\begin{figure*}[!htb]
\centerline{
\epsfig{figure=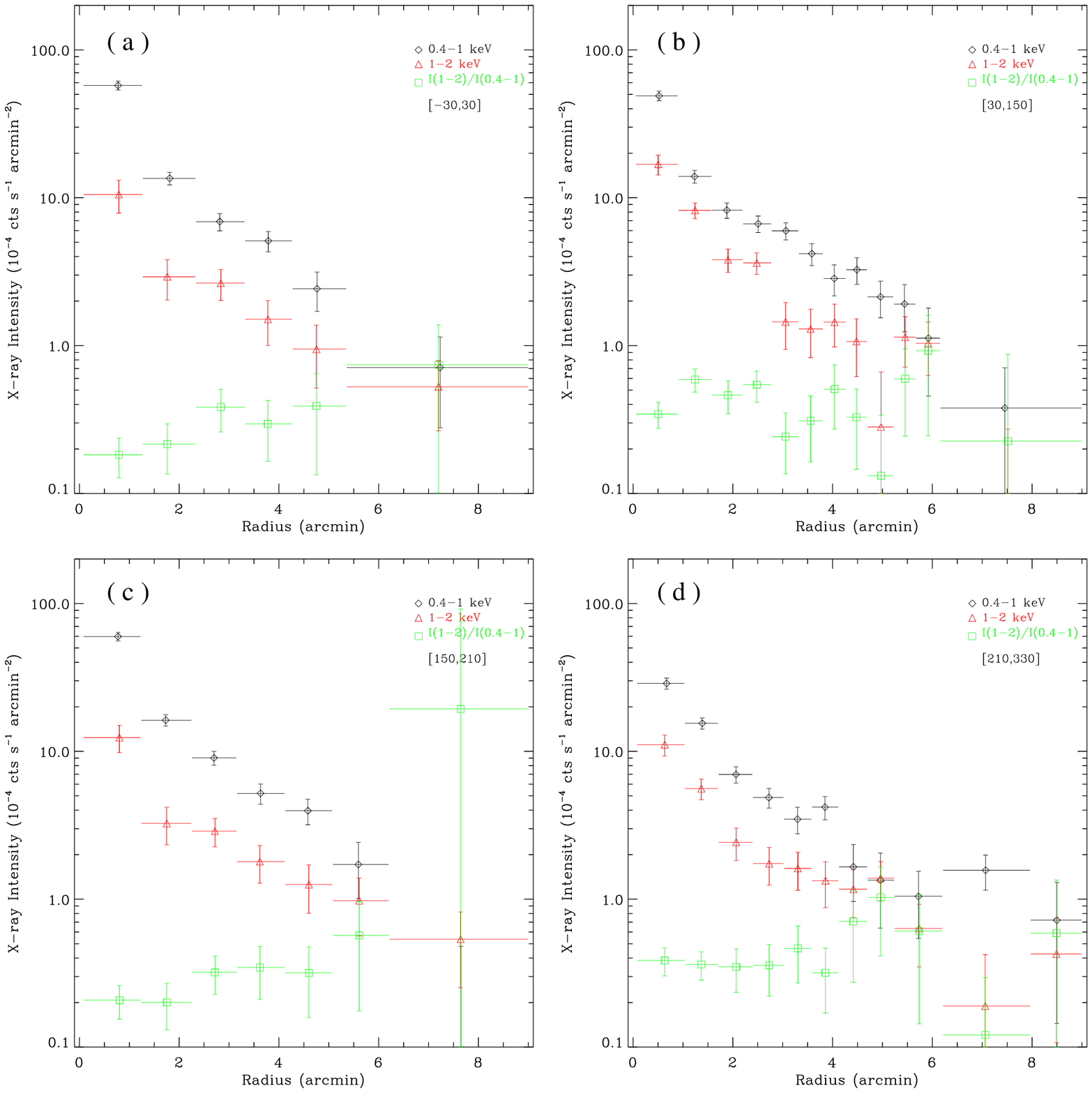,width=\textwidth,angle=0,clip=}
}
%\centerline{
%\epsfig{figure=M104_b2_rsb_dif_q3.ps,width=0.5\textwidth,angle=90,clip=} 
%\epsfig{figure=M104_b2_rsb_dif_q4.ps,width=0.5\textwidth,angle=90,clip=} 
%}
\caption{Azimuthally-averaged
radial intensity profiles of the diffuse emission in the 0.4-1 ({\sl
  black diamonds}) and 1-2 keV ({\sl red triangles}) bands. Also shown
are the intensity ratios between the two bands ({\sl green squares}). 
The azimuthal range, within which the profile is constructed, is
indicated in each panel. Angles are measured counterclockwise from
west, a direction assumed to be aligned with the major-axis, i.e., the
disk. The profiles are binned to achieve a signal-to-noise (with
respect to the combined stellar and background emission) better than
3 and a minimum of 900 (0.4-2 keV) counts per bin.
}
\label{fig:rsb_dif}
\end{figure*}

A map of the hardness ratio of the diffuse emission,
defined as $(I_{\rm 1-2~keV}-I_{\rm 0.4-1~keV})/(I_{\rm1-2~keV}+I_{\rm 0.4-1~keV})$,
is shown in Fig.~\ref{fig:dif}. 
%There is no significant correlation between the intensity and the hardness ratio. 
Interestingly, the emission appears softer along the disk region. 
%On the other hand, there is hint that the emission is hardest at regions
%$\sim$$2^{\prime}$ above the disk, roughly along the minor-axis. 
To quantify this trend, we examine 
radial intensity profiles for the 0.4-1 and 1-2 keV diffuse emission in four
quadrants, two of which sample the minor-axis and two along the major-axis. The profiles are shown in
Fig.~\ref{fig:rsb_dif}, along with the intensity ratio between the two
bands in corresponding radial intervals.
The two quadrants along the minor-axis clearly show softer emission in the
central $\sim$2\farcm5.

%\subsection{Substructures} {\label{subsec:spatial}}
%1 Examine the inner bulge/disk substructures of the diffuse emission:
%a soft component;

\subsection{Spectral properties of the hot gas} {\label{subsec:spec}}
The new {\sl Chandra} data allow us to carry out 
spatially-resolved spectral analysis of the diffuse emission.
To trace the low-intensity emission to large radii, it is crucial to
determine the local background,   
for which a ``double-subtraction'' procedure is used. 
Similar to the imaging analysis, a fiducial background is
first considered, which includes the quiescent instrumental background, the
PSF-scattered photons of the nucleus, the spatially non-uniform photons from 
unresolved sources, and the out-of-time events.  
In all the spectra analyzed below, such a fiducial background is
subtracted.

It is then necessary to characterize the sky background,
the spectra of which are extracted
from regions beyond $R=9^\prime$ from both ObsID~9532 and
9533. We excluded discrete sources and
avoided edges of the ACIS-I CCDs.  
We then use the FTOOL {\sl addspec} to coadd the spectra obtained from the
two observations and to generate averaged instrument responses
(rmf and arf). Although these spectra sample different sky
regions, such a procedure is warranted since the spectra
were extracted from similar detector regions (hence the instrument
response is similar).
To maximize the statistics, 
we also extracted spectra from the S2 CCDs of the two observations, excluding obvious
discrete sources, and similarly coadded the two spectra. 
The modeling of 
the sky background consists of three components. The first component is 
an absorbed PL, with a fixed photon index of 1.4 and a Galactic
foreground absorption column density (3.7$\times10^{20}{\rm~cm^{-2}}$),
representing the unresolved cosmic AGNs (Moretti et al. 2003).
The second component is a MEKAL model with a temperature of 0.1 keV, representing the emission
from the Local Bubble and the Galactic halo. As noted by LWH07, the
line of sight to Sombrero passes through the outskirt of the North Pole Spur (NPS;
Willingale et al.~2003), a
Galactic foreground soft X-ray-emitting feature. Hence a third
component, another MEKAL, is added to characterize the NPS emission. Absorption to
the two thermal components is allowed to vary below the Galactic
foreground value, and a temperature of 0.27$^{+0.04}_{-0.03}$ keV is found
for the NPS, consistent with the findings of LWH07 and Willingale et al.~(2003).
%The two sets of the merged spectra, i.e., from the I-chips and the S2 chip,
%are jointly fitted, with the normalizations of the two thermal
%components scaled accorinding to the underlying sky area. 
%The sky background spectra and the fitted model are shown in Fig.~\ref{fig:bkg_spec}.
In the following spectral fits, this model is adopted to account for the sky
background, with normalizations scaled to the underlying
sky area.

Guided by the analysis in \S~\ref{subsec:unr}, the diffuse emission
appears softer along the disk, but otherwise shows no significant
hardness variation in the outer regions. Hence we
first consider two distinct regions, the disk and the
bulge, chiefly to constrain the integrated spectral properties of the gas. 
The disk is defined by the two dashed sectors in
Fig.~\ref{fig:dif}, with inner-to-outer radii of 0\farcm5-2\farcm5 and azimuthal ranges of $\pm30^{\circ}$
from the minor-axis. A disk spectrum is extracted for each of the
three observations. 
%a disk spectrum is extracted from the upper (northern)
%half of this region, to minimize the effect of aborption by the dust lane. 
The bulge is defined as the region outside the disk and within 
a radius of 5$^\prime$, where the field of view is common to 
ObsIDs 9532 and 9533. A bulge spectrum is extracted from each of these
two observations. The disk and bulge spectra are then again coadded
respectively with {\sl addspec}. 

We first characterize the bulge emission. In addition to the sky
background, it is necessary to account for the stellar
contribution. We adopt the spectral model derived for
M32 to account for the CV+AB emission (MEKAL+MEKAL; \S~\ref{subsec:unr}), 
with the normalizations scaled to the K-band light enclosed in the
spectral extraction region. An additional PL, with a fixed
photon index of 1.7, is adopted to
account for the residual LMXB contribution. The normalization of this component is
allowed to vary and is effectively constrained by the hard band photons.
The stellar components are subject to the Galactic foreground absorption.
We then characterize the diffuse gas by another MEKAL model, again
subject to the Galactic foreground absorption. The fit is initiated by
allowing the abundance, as a single value for different metals, to vary.
An acceptable fit is obtained, giving a temperature of
0.58$^{+0.04}_{-0.04}$ keV and an abundance of 2.2$^{+1.6}_{-0.8}$, in the solar
abundance standard of Grevesse \& Sauval (1998). 
%The upper limit on
%the abundance, however, is not well constrained ($\lesssim5$).
To test possible metal-enrichment processes, we group
the metals into $\alpha$ elements (namely, C, N, O, Ne, Mg, Si, S, Ar
and Ca) and non-$\alpha$ elements (namely, Na, Al, Fe and Ni). 
The abundance in each group is linked. 
At a temperature of $\sim$0.6 keV,
the Fe L-shell complex, clearly present in the bulge
spectrum, is expected to have a dominating weight in constraining the
overall abundance.
Fixing the iron-like element
abundance at a value of twice solar, the best-fit gives an $\alpha$-to-Fe abundance ratio of 
 $Z_{\alpha}/Z_{Fe} = 0.5^{+0.5}_{-0.4}$, marginally indicating
that $\alpha$ elements are enriched to a lesser extent than iron,
as expected if metal-enrichment is primarily from Type Ia supernovae (SNe Ia).

A similar fitting procedure is carried out for the disk emission. 
A MEKAL model for the gas gives a best-fit temperature of
0.40$^{+0.10}_{-0.05}$ keV and a sub-solar (0.2$^{+0.4}_{-0.1}$) single abundance for all
metals. When the metals are divided into two groups, the fitted 
$Z_{\alpha}/Z_{Fe}$ is 0.8$^{+0.7}_{-0.6}$. While in the literature 
there have been arguments that a sub-solar abundance is unphysical given
the metal-enrichment by evolved stars and
is likely an artifact of fitting low spectral resolution CCD data to
multi-temperature gas (e.g., Buote et al.~2003; Kim \& Fabbiano 2003; David et al.~2010),
there remains the possibility that the hot gas near the Sombrero disk is
diluted by low-metallicity gas in the disk (see
\S~\ref{sec:disc}). The fitted disk temperature is statistically lower than
the bulge value, confirming the indication from hardness ratios (\S~\ref{subsec:unr}). 
The exact temperature of the disk-related gas can be even lower, as the
disk spectrum inevitably includes the harder bulge emission
from in front of the disk. To account for this effect, we add another
MEKAL component to fit the disk spectrum, whose temperature and
abundance are fixed
at 0.58 keV and 2 solar, respectively, to mimic the bulge emission. This
results in a best-fit temperature of 0.33$^{+0.13}_{-0.12}$ keV
and an abundance of $\sim$0.2 solar for the disk component.
The fit results for the bulge and disk spectra are summarized in Table \ref{tab:spec}.

\begin{deluxetable}{cccccc}
\tabletypesize{\scriptsize}
\tablewidth{0pt}
\tablecaption{Fits to the bulge and disk spectra$^a$}
\tablehead{
& \multicolumn{2}{c}{\underline{~~~~~~~~~~~~~~~~Bulge~~~~~~~~~~~}} &
\multicolumn{3}{c}{\underline{~~~~~~~~~~~~~~~~~~~~~~Disk~~~~~~~~~~~~~~~~~~~~~~~~~}} \\
Parameter & Mekal & FeMekal$^b$ &
 Mekal& FeMekal$^b$ & 2Mekal$^c$\\
}
\startdata
$\chi^2/d.o.f.$ & 68.4/84  & 66.1/84  &   53.2/53   & 52.9/53  & 51.5/52  \\
$T$ (keV) & 0.58$^{+0.04}_{-0.03}$  & 0.58$^{+0.03}_{-0.02}$  & 0.40$^{+0.10}_{-0.05}$  & 0.39$^{+0.09}_{-0.06}$  & 0.33$^{+0.13}_{-0.12}$/0.58      \\
$Z$ ($Z_\odot$) & 2.2$^{+1.6}_{-0.8}$  & 0.5$^{+0.5}_{-0.4}$$^d$ & 0.2$^{+0.4}_{-0.1}$    &  0.8$^{+0.7}_{-0.6}$$^d$     & 0.1$^{+0.2}_{-0.1}$/2   \\
Norm ($10^{-5}$) & 4.4$^{+3.8}_{-1.4}$  & 4.9$^{+0.4}_{-0.5}$  & 9.1$^{+6.2}_{-1.7}$  & 9.1$^{+0.8}_{-0.5}$    &  9.6$^{+25.5}_{-9.5}$/0.4$^{+0.5}_{-0.3}$ \\
Flux$^e$ & 18.5  & 16.8  & 5.1  & 5.0  & 3.6/1.5  \\
\enddata
\tablecomments{$^a$Only the model component for the diffuse gas is
  shown. See text for details on the sky background and stellar model
  components; $^b$A Mekal model in which the abundance of
  $\alpha$-elements is fit in terms of ratio to the iron abundance
  that is fixed at 2 solar;
  $^c$Two Mekal models; 
$^d$For the
  $\alpha$-to-Fe abundance ratio; $^e$0.3-2 keV unabsorbed flux in units of 10$^{-14}{\rm~ergs~s^{-1}~cm^{-2}}$.
}
\label{tab:spec}
\end{deluxetable}

%\begin{figure*}[!htb]
%\centerline{
%\epsfig{figure=bkg4_S2.ps,width=0.5\textwidth,angle=270,clip=}
%}
%\caption{Sky background spectra extracted from the I-chips ({\sl
%  black}) and the S2-chip ({\sl red}). The spectra are adaptively
%  binned to achieve a signal-to-noise better than 4 and a minimum of
%  30 counts per bin. The solid histograms show the
%  best-fit model. The three distinct model components are also shown
%  in broken histograms for the I-chips spectrum. Note the relatively
%  higher contribution of the power-law component in the S2-chip
%  specrtrum, which reflects the greater contribution of unresolved cosmic AGNs
%  in these regions at large off-axis angels. See text for details. }
%\label{fig:bkg_spec}
%\end{figure*}

In what follows we characterize the thermal properties of the bulge gas
using a spectral deprojection procedure. Bulge spectra are extracted
from consecutive sectors, centering at the nucleus, with inner-to-outer radii of
0$^\prime$-0\farcm5, 0\farcm5-1$^\prime$, 1$^\prime$-2$^\prime$,
2$^\prime$-3$^\prime$, 3$^\prime$-4$^\prime$, 4$^\prime$-5$^\prime$,
5$^\prime$-6$^\prime$ and 6$^\prime$-9$^\prime$. For sectors within
4$^\prime$, the azimuthal range is chosen to be $\pm60^\circ$ from
the northern minor-axis (dotted regions in Fig.~\ref{fig:dif}). 
These inner sectors are common to the field
of view of all three observations. Hence a spectrum is extracted
from each observation for each of these sectors, and spectra from the
same sectors are coadded using {\sl addspec}. Sectors outside 4$^\prime$
are essentially beyond the field of view of ObsID~1586, hence the
correspondingly spectra are extracted from the other two
observations. Paying attention to avoid CCD edges, we maximize the azimuthal range for these
outer sectors, as the diffuse emission shows no significant azimuthal
variation beyond 4$^\prime$, except in the outermost regions (Fig.~\ref{fig:rsb_dif_4q}). 

\begin{figure*}[!htb]
\centerline{
\epsfig{figure=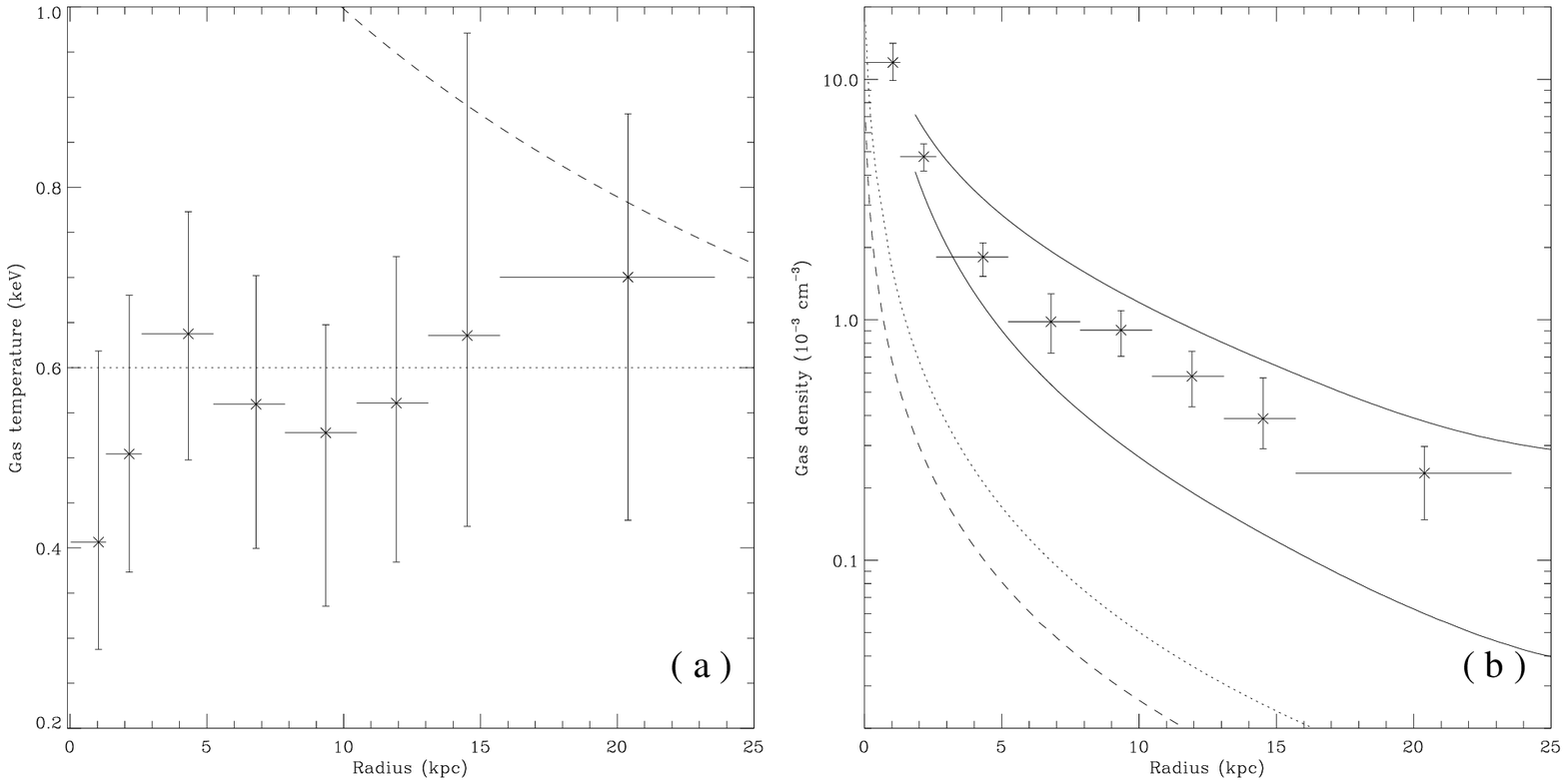,width=\textwidth,angle=0,clip=}
}
\caption{Radial distribution of gas temperature (a) and
 density (b) derived from spectral deprojection.
 The paired solid curves are the
 lower and upper bounds of the expected density distributions of an
 isothermal halo (with a fixed gas temperature of 0.6 keV) 
in hydrostatic equilibrium, based on the
 gravitational mass distribution of Bridges et al.~(2007). The dashed curve
represents the prediction of a steady, spherical wind with a specific heat ratio
$\gamma=5/3$, while the dotted curve is for an isothermal wind, i.e., $\gamma$=1. See the Appendix for details.}
\label{fig:deproj}
\end{figure*}

We then apply the {\sl projct} model in XSPEC to spectral
deprojection. We account for the fact that only a fraction of the area
within a
given annulus is covered, due to the selected azimuthal range and
source removal. A MEKAL model is adopted to characterize the diffuse
emission, while the sky background and stellar emission are accounted
for in the same way as for the above integrated bulge and disk spectra.
Our primary goal here is to derive radial distributions of the gas
temperature and density. Unfortunately the limited counting statistics do not allow us to examine a
radial variation of the abundance. Hence we fix the abundance at 2 solar but allow
the temperature to vary among different sectors. The resultant fit is
found to be an adequate characterization of all the spectra. The derived 
radial distributions of gas temperature and density are shown in Fig.~\ref{fig:deproj}.
The gas density has been converted from the MEKAL normalization,
assuming a volume filling factor of unity. It is noteworthy that the
derived gas density roughly inversely scales with the square root of
the presumed abundance, since the spectrum of a $\sim$0.6 keV gas is 
dominated by metal lines as long as the the abundance is near- or
super-solar.

\section{Discussion} {\label{sec:disc}}
Based on new {\sl Chandra} data, we have confirmed LWH07's detection of diffuse
hot gas in and around the bulge of Sombrero. A key question to address
is the origin and energetics of this gas. In principle, hot gas in and around
galaxies can originate either
externally from accretion of the IGM, or internally
from stellar feedback processes such as stellar winds heated by supernova
explosions. Feedback from an active nucleus, when it exists, can
provide additional heating to the gas. We examine these possibilities
in the following.

%\subsection{An extertal origin of the hot gas?} 
A generic prediction from current theories of galaxy formation is the presence of hot
gaseous halos around present-day disk galaxies, which are the relic
of accreted IGM heated to X-ray-emitting temperatures,
chiefly by accretion shocks and gravitational compression (White \& Rees 1978; White \& Frenk 1991).
% Toft et al.~2002; Rasmussen et al.~2009).
%Near the center of a disk galaxy, the cooling of 
%the hot gas is enhanced because of increased gas density. 
The development of such hot gaseous halos has been the subject of
various recent hydrodynamical simulations (e.g., Toft et al.~2002; 
Rasmussen et al.~2009; Crain et al.~2010).
These simulations generally find that the X-ray luminosity is a strong function of the gravitational mass of the host galaxy,
as characterized by its circular rotation speed ($L_X \propto V_c^{5-7}$).
With a maximum circular rotation speed  $V_c\approx370{\rm~km~s^{-1}}$ (Bajaja et al.~1984), 
Sombrero is among the most massive disk galaxies in the local universe
and hence may provide a strong constraint to this specific prediction of galaxy formation theories. 
LWH07 found that the amount of hot gas 
in the bulge/halo of Sombrero, with a measured X-ray luminosity of $\sim$3$\times10^{39}{\rm~ergs~s^{-1}}$,  is more than an order of magnitude lower than 
the value predicted by the simulations of Toft et al.~(2002). 
Recently, Rasmussen et al.~(2009) updated the Toft et al.~simulations using a refined, higher-resolution treatment. They
found that the 
detected amount of hot gas in Sombrero is comparable to their model 
prediction for a galaxy this massive. 
In an independent set of simulations by Crain et al.~(2010), the
predicted X-ray luminosities for Sombrero-like massive disk galaxies
show a substantial scatter, whose lower bound is 
in agreement with the LWH07 measurement.
From the above spectral analysis, we update the total 0.3-2 keV diffuse
X-ray luminosity to be $\sim$2.0$\times10^{39}{\rm~ergs~s^{-1}}$.
Hence an external origin of the hot gas in Sombrero cannot be ruled
out on the basis of integrated X-ray luminosity. 
However, current numerical models typically lack a
comprehensive treatment of internal feedback processes,
such as nuclear and stellar feedback, which are expected to regulate
the hot gaseous halo both mechanically and chemically.
For instance, none of the Rasmussen et al.~and Crain et
al.~simulations include nuclear feedback; these simulations
implement the chemical feedback from SNe Ia, but either neglect or
oversimplify their mechanical feedback. 
Nevertheless, the large population of old stars in Sombrero 
implies that SNe Ia and stellar mass-loss play an important role in
shaping the observed diffuse X-ray emission. One such piece of evidence is the super-solar
metal abundance inferred from the spectral fit (\S~\ref{subsec:spec}), which argues against
the accreted IGM as a dominant origin of the bulge hot gas.
Below we shall consider the internal feedback processes in Sombrero.

It is not trivial to trace and quantify the mechanical feedback from galactic nuclei. 
In particular, for those ``inactive''
nuclei showing a low radiation efficiency, much of the nuclear feedback is expected to be carried out (and/or
accompanied) by jets of relativistic particles (e.g., Yuan et al.~2009).
Allen et al.~(2006) found an empirical relation between accretion rate
and jet power in X-ray-bright elliptical galaxies, quantitatively as: ${\rm log}(L_{\rm Bondi}/10^{43}{\rm~ergs~s^{-1}}) 
= 0.65 + 0.77 {\rm log}(L_{\rm jet}/10^{43}{\rm~ergs~s^{-1}})$. 
The Bondi accretion rate of their sample nuclei
ranges from $3.5\times10^{-4}{\rm~M_\odot~yr^{-1}}$ to $4.6\times10^{-2}{\rm~M_\odot~yr^{-1}}$, 
encompassing the value estimated for the nucleus in Sombrero (\S~\ref{subsec:circum}).
This relation would imply a potential jet power of $\sim$$1.4\times10^{42}{\rm~ergs~s^{-1}}$ for Sombrero.
In principle, this powerful nuclear feedback can result in cavities in the hot gaseous halo, 
especially along the direction of the kpc-scale linear radio structure found by Gallimore et al.~(2006), but no such evidence is seen in our X-ray images. 
If the radio structure is truly jet-powered, an equipartition magnetic field strength of $B_{\rm eq}\approx $ 4.5 ${\mu}$G can be inferred. 
To estimate this value, we have adopted the measured flux density of 3.1 mJy and length of 54$^{\prime\prime}$ at 5 GHz (Gallimore et al.~2006) and assumed a canonical jet spectral index of 0.7 for the radio jet and a line-of-sight depth of 15$^{\prime\prime}$ (limited by the image resolution). 
The corresponding magnetic field pressure, $\sim$$8\times10^{-13}{\rm
  ~dyn~cm^{-2}}$ is $\sim$10 times smaller than the thermal pressure of hot gas in the inner kpc ($\sim$$8\times10^{-12}{\rm~dyn~cm^{-2}}$), but becomes comparable with the latter at radii beyond $\sim$10 kpc. 
Interestingly, Fig.~\ref{fig:unr}d and Fig.~\ref{fig:rsb_dif_4q}
suggest an elongation along the direction of the radio structure
(i.e., toward southeast) in the diffuse X-ray emission which otherwise
appears rather symmetric. This may be the site where the assumed jet becomes dynamically important to the hot gas.    
The mechanical energy deposited by the jet can be characterized by the $PV$ work, 
$\sim$$2.2\times10^{53}{\rm~ergs}$, if we adopt the thermal pressure instead of the smaller equipartition magnetic field pressure. 
The timescale for this energy to be consumed by the gas is rather uncertain. 
If we assume that the jet grows to the observed size at the sound speed of the hot gas, 
we estimate a timescale of 2.8 Myr and hence an average mechanical
heating rate of 1.2$\times10^{39}{\rm~ergs~s^{-1}}$. 
Similarly, if the circumnuclear X-ray/optical line feature
(\S~\ref{subsec:circum}) is powered by the jet, a heating rate of
$\sim$$10^{39}{\rm~ergs~s^{-1}}$ can be inferred.
These heating rates, while being comparable to the observed X-ray
luminosity of the hot gas, are much less than the integrated heating
rate by SNe Ia, as we shall show below. 
They also suggest a rather inefficient extraction of the
possible jet power. In principle, the nuclear feedback may cause
distinct effects in the circumnuclear ISM, e.g., a steep pressure
gradient (e.g., Bland-Hawthorn 1995). At present, probing such an effect 
is however hampered by the bright X-ray nucleus and the lack of suited
optical spectroscopic observations.

%\subsection{Stellar feedback}
The most violent form of stellar feedback is related to massive stars
(cf.~Veilleux, Cecil \& Bland-Hawthorn 2005).
Hot gas heated by stellar winds and core-collapse Type II SNe is commonly detected
in active star-forming galaxies with an edge-on disk perspective
(e.g., Strickland et al.~2004; Tullmann et al.~2006; Li et al.~2008).  
Low level star formation is likely present in the disk of
Sombrero. H$\alpha$ emission and radio continuum emission are
detected along the disk (LWH07; Bajaja et al.~1988), both implying a star
formation rate (SFR) of $\sim$$0.1-0.2{\rm~M_\odot~yr^{-1}}$. 
Using the $L_X$-SFR relation found by Strickland et al.~(2004) and Li
et al.~(2008), a 0.3-2 keV luminosity of (2-8)$\times10^{37}{\rm~ergs~s^{-1}}$
is expected. Compared to the measured diffuse X-ray luminosities  
along the disk and in the bulge (Table \ref{tab:spec}), we conclude that star-forming
activities contribute little to the detected diffuse X-ray emission in
Sombrero, although their mechanical input to the hot gas cannot be ruled out.

Lastly, we consider feedback from old stellar populations.
Empirically, the old stars in a galactic bulge continuously
deposit energy and mass
into the interstellar medium (ISM)
at rates of $\sim$$1.1\times10^{40}$$[L_K/(10^{10} L_{K,\odot})]$
${\rm~ergs~s^{-1}}$ and $\sim$$0.02$$[L_K/(10^{10} L_{K,\odot})]$
${\rm~M_\odot~yr^{-1}}$
(e.g., Mannucci et al. 2005; Knapp, Gunn \& Wynn-Williams 1992; see also Appendix), respectively,
where $L_K$ is the K-band luminosity of the bulge.
A ``missing stellar feedback'' problem is often found in gas-poor, i.e., low $L_X/L_K$, early-type galaxies,
where the X-ray luminosity, mass and metal content of the hot gas inferred
from observations account for only a small fraction of the expected stellar feedback
(e.g., O'Sullivan, Ponman \& Collins 2003; David et al.~2006; LWH07;
Li \& Wang 2007; Wang 2010).
Sombrero, with $L_X/L_K\sim 10^{28}{\rm~ergs~s^{-1}}/10^{10}
  L_{K,\odot}$, shows such a discrepancy.
From the spectral analysis                                                                                
we have measured $L_X\approx2.0\times10^{39}{\rm~ergs~s^{-1}}$ and $M_{\rm
  gas}\approx2.9\times10^8{\rm M_{\odot}}$, whereas the expected SNe
Ia heating rate is
$\sim$$2.2\times10^{41}{\rm~ergs~s^{-1}}$ and the expected mass input
rate is $\sim$$0.4{\rm~M_{\odot}~yr^{-1}}$ (taking only $\sim$$7\times10^8$ yr to accumulate
the observed amount of hot gas). We note that the mass input rate is
several times higher than the rate of on-going star formation on the disk,
and that the mass of cold gas in both the disk
and the bulge is no more than a few $10^8{\rm~M_\odot}$ (Bajaja et al.~1984; Matsumura \& Seki 1989; after rescaling the adopted distances therein),
together indicating that the bulk of mass returned by the old stars is unlikely 
to be stored within where it is produced.
Naturally, the energy and mass discrepancies can be explained with the presence of an outflow of hot gas, through which the ``missing'' energy and mass
are transported outside the regions covered by the observations.
Such an outflow is expected to show distinct thermal structures, i.e., the radial distributions of gas temperature and density.

Indeed, with its diffuse X-ray emission being traced to at least $\sim$23 kpc from the center, 
Sombrero is an ideal target for examining the thermal structures. The simplest case is that the hot gas is in hydrostatic equilibrium with the gravitational potential. 
The predicted density distribution for an isothermal gas halo, as implied by the nearly constant gas temperature, is shown as solid curves in Fig.~\ref{fig:deproj}. 
The gravitational potential distribution is obtained by modeling the
observed kinematics of globular clusters out to a radius of $\sim$25 kpc (Bridges et al.~2007).
Within the current uncertainties in both the X-ray and kinematic data, the observed density distribution is consistent with a hydrostatic equilibrium case. 
However, as stars would continuously return mass to the interstellar
space, a hydrostatic gas halo is expected to increase its mass at the
rate of integrated mass input given above, which is not
observed. We note that the gas mass fraction, with respect to the 
gravitational mass, is only $\sim$0.3\% within a radius of 25 kpc.
Again, this indicates that Sombrero is losing its hot gas.

To examine the gas temperature and density distributions regulated by 
feedback from old stellar populations, we construct
a simple one-dimensional hydrodynamic model (cf. Appendix for details)
for a galactic-scale transonic wind.   
The results are shown as dashed and dotted curves in Fig.~\ref{fig:deproj}.  
Clearly, the wind density drops rapidly as the result of outward
expansion, being $\sim$10 times lower than the observed value at radii
beyond 10 kpc. The corresponding
X-ray luminosity is more than 100 times lower than the observed value.
We note that the adopted spherical symmetry in our model is
practically simplified, in view of Sombrero's starlight distribution. 
The shape of the gravitational potential in the inner few arcminutes, in
particular, is regulated by the ellipsoidal bulge and the disk.
Moreover, gas returned by the evolved stars may carry 
a net angular momentum due to the rotation of the bulge.
The flattening and rotation of stars may conspire to
modify the dynamics of the hot gas and hence its morphology (e.g.,
Ciotti \& Pellegrini 1996; D'Ercole \& Ciotti 1998).
This is hinted at by the boxy morphology of the diffuse emission in the
inner regions (Fig.~\ref{fig:unr}).
 Nevertheless, the importance of such an effect is expected to decrease
with radius, and any resultant increase in the amount of hot gas is unlikely
sufficient to account for the observed discrepancy by a factor of
$\sim$100, without producing an even greater amount of cold gas in the
disk (D'Ercole \& Ciotti 1998).
 A detailed modeling of a flattened, rotating outflow is
beyond the scope of the present work.

An alternative, qualitatively viable solution is that the hot gas forms
a subsonic outflow (e.g., David, Forman \& Jones 1991; Tang et al.~2009a), the thermal
structure of which still resembles, and hence is difficult to
distinguish from, a quasi-hydrostatic halo, especially at large radii.
This can be seen by estimating the asymptotic flow velocity $V_g=\dot{M}/(4{\pi}R^2{\mu}{m_H}{n_g})$. 
Adopting the measured gas density ($n_g$) at the outermost radius
($\sim$20 kpc; Fig.~\ref{fig:deproj}) and the total mass outflowing
rate ($\dot{M}$) of 0.4${\rm~M_{\odot}~yr^{-1}}$ (i.e., the collective
mass loss from the old stars), we have
$V_g\approx20{\rm~km~s^{-1}}$, which is much less than the local sound speed.
Recent numerical simulations of bulge outflows indicate that the
thermal structures are time-dependent and sensitively depend on the
formation history and environment of the host galaxy (Tang et
al.~2009a).
Interestingly, the simulations predict that the effective temperature
 of the outflow is relatively low in the central region, chiefly due
 to the non-uniform 
 thermalization of the SNe Ia energy in a dense environment (Tang et al.~2009b). The 
 low gas temperature near the disk of Sombrero seems to be consistent
  with this prediction. In addition, thermal conduction and/or charge exchange between
  the hot gas and the cold gas residing in the disk may
  lead to softer X-ray emission, a situation suggested to be present in the
  circumnuclear region of M31 (Li, Wang \& Wakker 2009; Liu et al.~2010).
  
%A hint of this kind is the apprently lower intensities seen along the major-axis, compared to those along the minor-axis in the similar radial ranges (within $\sim$4$^\prime$).
Another possibility is that the observed soft X-rays arise from the
interfaces between a very hot (a few keV) wind and an entrained, cold
medium (see e.g., Veilleux et al.~2005). In this case, the wind
temperature would be consistent with the empirical 
specific energy input per gas particle from evolved stars. However,
the current X-ray data of Sombrero 
provide no compelling evidence for such a wind, e.g., in terms of
diffuse hard X-ray emission and/or iron K-shell lines as observed in
the superwind galaxy M82 (Strickland \& Heckman 2007).
On the other hand, extraplanar cold gas
can be probed as extinction features against the bulge starlight
(e.g., Howk \& Savage 1997, 1999; Li et al.~2009). We find no such
evidence in {\sl HST} broad-band optical images of Sombrero.

\section{Summary} {\label{sec:sum}}
Based on new and archival {\sl Chandra} observations, we have studied
the X-ray nucleus and the diffuse hot gas in the bulge-dominated,
edge-on, Sa galaxy Sombrero. The study can be summarized as follows: 

1. The 0.3-8 keV luminosity of the nucleus appears to be
constant at $\sim$$2.4\times10^{40}{\rm~ergs~s^{-1}}$ on three epochs between December 1999
and April 2008, but drops by a factor of two in the November 2008 observation.
The observed X-ray luminosity is only one part in 10$^7$ of the {\sl
  Eddington} luminosity for the SMBH, characteristic of a radiatively
inefficient accretion flow. No significant spectral
variation is found associated with the flux change. The nuclear
spectrum can be characterized by a power-law with a
photon index of $\sim$2 and a moderate absorption column density of
$\sim$$3\times10^{21}{\rm~cm^{-2}}$. 

2. After properly accounting for the contribution of unresolved stellar
   sources, in particular CVs and ABs whose quasi-universal X-ray
   emissivity is recently established, we isolate the truly diffuse emission of
   hot gas. The extent of this diffuse emission is significantly larger
   than that of the star light, reaching a galactocentric radius of at least
   23 kpc, where it becomes indistinguishable from the local sky background.

3. The hot gas shows a temperature of $\sim$0.6 keV with little
statistically significant variation across the field of view, 
except around the disk where a lower temperature of $\sim$0.3
keV is hinted. This cooler gas may result from thermal conduction
and/or turbulent mixing with 
the cold ISM in the disk.
 
4. While the total intrinsic luminosity of the hot gas is comparable
   to the prediction by the latest galaxy formation models, in which
   a hot gaseous halo is a generic feature of a disk galaxy formed by 
   the accretion of the IGM, we show that the role of internal feedback
   processes, in particular the energy input by SNe Ia and the mass 
   input by evolved stars, cannot be neglected.

5. A simple hydrodynamic model characterizing the stellar feedback and
   the gravitational potential in Sombrero, in which the hot gas is in
   the form of a supersonic wind, however, fall far short of reproducing the
   observed amounts of X-ray luminosity and gas mass. While recent
   more sophisticated numerical simulations suggest that a subsonic
   outflow is a viable solution, understanding of the
   thermodynamical structure of the hot gas in Sombrero and in other
   typically intermediate-mass early-type galaxies continues to be pursued.

\vspace{0.2 cm}
ZL is grateful to the useful comments and discussions with Robert Crain, Larry David, Hui Dong and Dan Harris. This work is supported by the SAO grant G08-9088.

\appendix
\section{A Galactic Wind Model} {\label{set:wind}}
%Theoretical models for galactic-scale gas flows have long been developed
%(e.g., White \& Chevalier 1983, 1984; Loewenstein \& Mathews 1987; Ciotti et al.~1991),
%most of which are in the scope of elliptical galaxies for which spherical symmetry is
%reasonable approximation. White \& Chevalier (1983) studied the steady state galactic wind,
%in which they have assumed constant supernova energy input and stellar mass loss, the rates of which are estimated
%at the current epoch. Stimulated by the fact that both the supernova rate
%and the stellar mass loss rate could
%be much higher at early epochs, Ciotti et al.~(1991) modeled the time-dependent evolution
%of galactic gas flows. They proposed that the gas flow would evolve through up to three consecutive stages:
%the wind, subsonic outflow and inflow phases, which can in principle account for the large
%scatter in the observed X-ray luminosities of galaxies with similar optical luminosities (e.g., Ellis \& O'Sullivan 2006).
%While some of these models succeeded in predicting the X-ray properties of gas in several aspects,
%such as the total X-ray luminosity, direct comparison with observations has been so far restricted in the scope
%of integrated properties of the gas.
%More suggestive comparison is to be carried out, in terms of
%confronting the model-predicted spectral and spatial properties of gas with current X-ray observations on sub-galactic
%scales. To do so, we construct a simple one-dimention steady-state galactic wind model, as follows.

Hydrodynamic models for galactic-scale flows have long been developed
(e.g., Mathews \& Baker 1971; White \& Chevalier 1983, 1984; Loewenstein \&
Mathews 1987; David, Forman \& Jones 1990; Ciotti et al.~1991),
most of which are in the scope of elliptical galaxies for which spherical symmetry is
a reasonable approximation. 
The essence of these models includes the characterization of  
the gravitational potential and 
the mass and energy deposition (i.e., stellar feedback) to the gas
flow. Such properties, which together determine the gas dynamics,
are practically difficult to constrain from direct observations.
Our approach is to construct a model with simple, but essential physical considerations, aiming to confront the modeled gas properties with observational results.

\subsection{Physical assumptions and formulation} {\label{sec:formulation}}
Our first assumption is that the gas flow is in a steady state and a spherical symmetry.
The dynamical timescale of a wind is much shorter than the timescale of
galaxy evolution, hence 
it is reasonable to assume that 
the mass and energy input from evolved stars, which together supply the gas flow, are at constant rates.
Spherical symmetry is not only a practical assumption, as used in 
most theoretical models, but also is supported in
the X-ray morphologies of galactic spheroids.
All physical parameters considered hereafter, e.g., gas density and temperature, 
are functions of galactocentric radius.
Second, we assume that the mass and energy input spatially follow the distribution of
stars. The mass input is essentially contributed by the ejecta of evolved stars, such as stellar winds and planetary nebulae;
the energy input is primarily provided by mechanical energy from Type Ia SNe, with additional orbital energy carried by the stellar ejecta. 

In galaxies where gas outflows exist, the X-ray luminosity is typically as low as only a few percent of the energy input. 
%The introduction of energy loss by radiative cooling is found to make
%barely change to the dynamics as determined for an adiabatic wind. 
Thus we neglect the effect of radiative cooling on the gas dynamics.
%an assumption by which we can largely reduce the computational expense. 
This is a good approximation as the cooling timescale is
much longer than the dynamical timescale, probably except for the central region. 
%We note that this assumption is not valid in the case of X-ray bright elliptical galaxies, 
%where gas inflow is believed to be prevailing. In that case,
%the cooling timescale is comparable to or even significantly shorter than 
%the dynamic timescale, making the effect of cooling on the dynamics crucial. 
Other assumptions that have been made include neglecting thermal conduction, viscosity,
and self-gravity of the gas. 

Now the physical properties of the gas can be described by the hydrodynamic equations:
\begin{equation}
  \frac{1}{r^2}\frac{d}{dr}({\rho}ur^2) = \dot{m}(r),
\label{eq:mass}
\end{equation}
\begin{equation}
  {\rho}u\frac{du}{dr} = -\frac{dP}{dr}-{\rho}g-\dot{m}(r)u,
\label{eq:momentum}
\end{equation}
\begin{equation}
  \frac{1}{r^2}\frac{d}{dr}
  [{\rho}ur^2(\frac{1}{2}u^2 +\frac{\gamma}{{\gamma}-1}\frac{P}{\rho})]
  = -{\rho}gu+\dot{E}(r),
\label{eq:energy}
\end{equation}
where $u$, $\rho$, and $P$ are the velocity, density and pressure of gas, respectively. 
$\gamma$ is the ratio of specific heats. 
g(r) is the gravitational force given by
\begin{eqnarray} \label{eq:potential}
  g(r) = \frac{G[M_s(r)+M_d(r)]}{r^2}, \\ 
  M_s(r) = \int_0^r 4{\pi}{\rho}_s(r)r^2 dr, \\ 
  M_d(r) = \int_0^r 4{\pi}{\rho}_d(r)r^2 dr,
\end{eqnarray}
where ${\rho}_s(r)$ and ${\rho}_d(r)$ are the density distributions of stellar mass 
and dark matter, respectively. $\dot{m}$(r) and $\dot{E}$(r) are the mass and energy 
input rates per unit volume, which are assumed to be proportional to the density of 
stellar mass ${\rho}_s$(r).

%Note that the above equations are applicable for both wind 
%and subsonic outflow. The solutions for the two cases distinguish
%with each other by different boundary conditions. Readers may refer to
%Appendix A for details.

\subsection{Galaxy modeling} {\label{sec:galaxy}}
A valid solution of the above hydrodynamic equations stands on a specific realization of 
the host galaxy, i.e., the spatial
distributions of stars and dark matter and the mass and energy input rates from stars.
%For the stellar distribution, we adopt the Hernquist profile (Hernquist 1990)
%\begin{equation}
%  \rho_s(r) = \frac{M_s}{2\pi}\frac{r_s}{r(r+r_s)^3},
%\label{eq:Henquist}
%\end{equation}
%where $r_s$ is the scale radius and $M_s$ is the total stellar mass.
%The Hernquist profile, when projected, closely approximates the de Vaucouleurs's law,
%which is empirically used to describe the stellar surface brightness profile of elliptical galaxies and bulges (Hernquist 1990).
%We note that this profile predicts a cusp at the galactic center. Since the mass input
%follows the stellar distribution, the resultant central gas density is infinite. In reality, the existence of discrete X-ray sources almost always prevent us
%from directly probing the diffuse emission of gas near the very center of the host galaxy,
%therefore for simplicity we save our effort in relaxing a central cusp
%in the stellar distribution. 
To derive the spatial distribution of stars, we deproject a de Vaucouleurs's law, 
which is empirically used to describe the stellar surface brightness profile of elliptical galaxies and bulges. We adopt an effective radius $R_e$ = 105$^{\prime\prime}$ ($\sim$4.6 kpc) for the bulge of Sombrero (Bendo et al.~2006). 
We note that the stellar disk contributes less than 20\% of the total stellar mass in Sombrero and is neglected. 
The total stellar mass $M_{st}$, 1.5$\times10^{11}{\rm~M_{\odot}}$, is determined from the color-dependent mass-to-light ratio of Bell et al.~(2003) based on the 2MASS K-band luminosity of Sombrero.
For the dark matter halo, we adopt the NFW profile (Navarro, Frenk and White 1996), 
%in which density decreases following $r^{-3}$ at large radii,
\begin{equation}
  \rho_d(r) = \frac{\rho_{d0}}{{\frac{r}{r_d}(1+\frac{r}{r_d})^2}}.
\label{eq:NFW}
\end{equation}
$\rho_{d0}$ can be replaced by an algebraic combination of the total dark matter mass $M_{dt}$ and the concentration parameter C${\equiv}R_v/r_d$.
The galaxy boundary is defined to be the virial radius, $R_v$, which
scales with the cube root of the virial mass (total mass of stars and dark matter). 
Since we have good observational constraints on the distributions 
of stellar mass (based on K-band light) and gravitational mass (based
on kinematics of globular clusters out to $r$=25 kpc; Bridges et al.~2007), we fit the NFW profile to the difference between
the gravitational mass and the stellar mass to obtain $r_d$=27 kpc,
$R_v$=483 kpc and M$_{dt}$=6.8$\times10^{12}{\rm~M_{\odot}}$.

To complete the modeling,
we also need to specify the total stellar mass loss rate $\dot{m}_0$ and 
the collective energy input rate $\dot{E}_0$,
so as to determine $\dot{m}(r)$ and $\dot{E}(r)$ (both assumed to be proportional to $\rho_s$) via relations
\begin{equation}
  \int_0^{R_v}  4{\pi}r^2\dot{m}(r)dr = \dot{m}_0,
\label{eq:mass rate}
\end{equation}
\begin{equation}
  \int_0^{R_v} 4{\pi}r^2\dot{E}(r)dr = \dot{E}_0.
\label{eq:enery rate}
\end{equation}
$\dot{m}_0$ and $\dot{E}_0$ are the two key parameters in our model. 
%In case that the model is a good approximation of the actual gas dynamics, we expect that the observed X-ray emission of the gas be well fitted by the model predictions (see below), from which we then have constraints on $\dot{m}_0$ and $\dot{E}_0$.

For a given galaxy, there are also empirical methods to estimate $\dot{m}_0$ and $\dot{E}_0$.
Knapp, Gunn and Wynn-Williams (1992) suggested the following relation as a direct
measurement of the current mass
loss rate from evolved stars in elliptical galaxies, using the 2.2 ${\mu}m$ flux density
\begin{equation}
\dot{m} = 8\times10^{-4}(\frac{D}{\rm Mpc})^2(\frac{S_{2.2}}{\rm Jy}){\rm~M_{\odot}~yr^{-1}} = 0.02\times10^{-2}(\frac{L_K}{10^{10}L_{\odot,K}}){\rm~M_{\odot}~yr^{-1}}.
\label{eq:m0}
\end{equation}

%Ciotti et al. (1991) modelled the stellar mass loss rate by considering the evolution
%of a coeval stellar population. They gave
%\begin{equation}
%\dot{m} = 0.15(\frac{L_B}{10^{10}L_{\odot,B}})t_{15}^{-1.3}M_{\odot}yr^{-1}.
%\label{eq:m02}
%\end{equation}
%where $t_{15}$ is the age of the galaxy in units of 15 Gyr.

The energy input consists of that from Type Ia SNe and that from stars,
i.e., $\dot{E}_0 = \dot{E}_{0,{\rm SN}}+\dot{E}_{0,{\rm star}}$. 
By assuming a typical released energy of $10^{51}{\rm~ergs}$ for a single SN explosion, 
\begin{equation}
\dot{E}_{0,{\rm SN}} = 1.1{\times}10^{40}(\frac{R_{{\rm SN}}}{0.035{\rm SNuK}})(\frac{L_K}{10^{10}L_{\odot,K}}){\rm~ergs~s^{-1}},
\label{eq:E0}
\end{equation}
where $R_{\rm SN}$ is the
observed Type Ia SNe rate in units of SNuK (one per $10^{10}L_{\odot,K}$ per 100 yr) for
early-type galaxies at low redshifts (Mannucci 2005).

The stellar ejecta also carry mechanical energy obtained from the orbital motion of the progenitor star,
\begin{equation}
\dot{E}_{0,{\rm star}} = \frac{1}{2}\dot{m}_0{\sigma}^2 \approx 0.6{\times}10^{39}(\frac{L_K}{10^{10}L_{\odot,K}})(\frac{\sigma}{300{\rm~km~s^{-1}}})^2{\rm~ergs~s^{-1}}, 
\label{eq:E_s}
\end{equation}
where $\sigma$ is the stellar velocity dispersion. Clearly, $\dot{E}_{0,{\rm SN}}$ is the
dominant form of energy input.

Closely relevant is the metal-enrichment of gas predominantly by the SNe.
We take the iron enrichment as an example. Nomoto, Thielemann and Yokoi (1984) calculated the Fe yield per
Type Ia SN to be 0.7$M_{\odot}$, about half of the total released mass. 
Assuming a complete mixture of the iron atoms with the gas, the iron abundance can be estimated as
\begin{equation}
  Z_{\rm Fe}=Z_{\rm Fe,star}+9.7(\frac{3.16{\times}10^{-5}}{[n_{\rm
  Fe}/n_{\rm H}]_{\odot}})(\frac{R_{{\rm SN}}}{0.035 {\rm SNuK}}),
\label{eq:Z_Fe}
\end{equation}
which is independent of the total stellar content.
According to Grevesse \& Sauval (1998), $[n_{{\rm Fe}}/n_{\rm H}]_{\odot}=3.16\times10^{-5}$, 
thus by assuming that $Z_{\rm Fe,star}$ equals solar, Eq.(\ref{eq:Z_Fe}) gives
$Z_{\rm Fe,gas}=10.7$. 
%Alternatively, a solar abundance standard of $[n_{Fe}/n_H]_{\odot}=4.68\times10^{-5}$
%by Anders \& Grevesse (1989) gives $Z_{Fe,gas}=6.0$.

%The above estimates of $\dot{m}_0$ and $\dot{E}_0$, however, are averaged values obtained from a large sample
%of galaxies. The actual values for individial galaxies, while cannot be easily probed from
%direction observations, may deviate from the average significantly. Our aim of confronting the model with X-ray observations provides an alternative way to constrain the values of $\dot{m}_0$ and $\dot{E}_0$, such that the resultant model best fits the observed gas properties (see below).

\subsection{Solution} {\label{subsec:solve_wind}}
To solve Eqs.(\ref{eq:mass}) - (\ref{eq:energy}), 
three boundary conditions, or three equivalent constraints on the variables, are generally needed. 
Given a natural assumption that the gas velocity is zero at the center,
we have the first boundary condition,
\begin{equation}
   u{\arrowvert}_{r=0} = 0.
\label{eq:u0}
\end{equation}

With a further assumption that the derivatives of velocity and temperature
are finite at the center, Eq.(\ref{eq:energy}) and (\ref{eq:u0}) require that

\begin{equation}
   kT_0 \equiv kT{\arrowvert}_{r=0} = {\mu}_g m_{\rm H}\frac{P}{\rho}{\arrowvert}_{r=0} = {\mu}_gm_{\rm H}\frac{\gamma-1}{\gamma}\frac{\dot{E}_0}{\dot{m}_0},
\label{eq:T0}
\end{equation}
where ${\mu}_g$ is the mean molecular weight of gas taken to be 0.6,
$m_{\rm H}$ and $k$ have the conventional meanings.

Now it is convenient to introduce a dimensionless variable,
the Mach number $M \equiv u/c_s$, where ${c_s}\equiv({\gamma}P/{\rho})^{1/2}$ is the sound speed of gas. 
With Eq.(\ref{eq:u0}) and (\ref{eq:T0}), one can show that equations
(\ref{eq:mass})-(\ref{eq:energy}) reduce to
a first order differential equation of the Mach number
\begin{eqnarray}
\label{eq:Machdiff}
  \frac{dM^2}{dr} = \frac{M^2}{M^2-1}(1+\frac{{\gamma}-1}{2}M^2)g(r,M), \\
  g(r,M^2) = \frac{4}{r}-\frac{{\gamma}+1}{{\gamma}-1}\frac{\frac{dW}{dr}}{L-W} \nonumber
             -(1+{\gamma}M^2)[\frac{\frac{dF}{dr}}{F}+\frac{\frac{dL}{dr}}{L-W}], \nonumber
\end{eqnarray}
where
\begin{equation}
  F(r) = \int_0^r 4\pi\dot{m}(r)r^2dr,
\label{eq:F}
\end{equation}
\begin{equation}
  L(r) = \int_0^r 4\pi\dot{E}(r)r^2dr,
\label{eq:L}
\end{equation}
\begin{equation}
  W(r) = \int_0^r F(r)\frac{d{\phi}}{dr}dr.
\label{eq:W}
\end{equation}

In general the right-hand side of Eq.(\ref{eq:Machdiff}) is singular when $M = 1$.
Correspondingly, the radius $r=r_c$ at which $M=1$ is called the {\sl sonic radius} or
{\sl critical radius}. 
A wind solution requires that the gas flow smoothly passes through the sonic radius.
This provides a third boundary condition to the solution,
\begin{equation}
   g(r,M){\arrowvert}_{r_c} = 0,
\label{eq:r_c}
\end{equation}
so that the right-hand side of Eq.(\ref{eq:Machdiff}) remains regular when $M=1$.

To derive the wind solution, the location of the sonic radius is first found by solving 
Eq.(\ref{eq:r_c}), which is simply an algebraic equation of $r$. 
Then {\sl L'Hospita}'s rule is applied to obtain the derivative of $M$ at $r_c$. 
Finally, a wind solution is found by integrating Eq.(\ref{eq:Machdiff}) starting from $r_c$ inward to
the center and from outward to the virial radius.
An adaptively stepping fifth order Runge-Kutta method is used when performing
 the numerical solution. 
%Fig.~\ref{fig:wind} shows a representative solution of the wind model for M104.

\end{document}